\documentclass[11pt,a4paper]{article}
\usepackage{jheppub}
\usepackage{epsfig}
\usepackage{amsmath}
\usepackage{latexsym}
\usepackage{amssymb}
\usepackage{dsfont}
\usepackage{color} 
\newcommand{\ud}{\mathrm{d}}

\newcommand{\uM}{\mathcal{M}}

\newcommand{\uTr}{\mathrm{Tr}}

\newlength\savedwidth
\newcommand\whline{\noalign{\global\savedwidth\arrayrulewidth
\global\arrayrulewidth 1pt}%
\hline
\noalign{\global\arrayrulewidth\savedwidth}}

\title{Unified framework for generalized and transverse-momentum dependent parton distributions within a 3Q light-cone picture of the nucleon}

\author[a]{C. Lorc\'e,}
\author[b]{B. Pasquini,}
\author[a]{and M. Vanderhaeghen}

\affiliation[a]{Institut f{\"u}r Kernphysik, Johannes Gutenberg-Universit{\"a}t,\\ Mainz, D55099, Germany}
\affiliation[b]{Dipartimento di Fisica Nucleare e Teorica, Universit\`a degli Studi di Pavia, \\and INFN, Sezione di Pavia, I-27100 Pavia, Italy}
\emailAdd{lorce@kph.uni-mainz.de}
\emailAdd{pasquini@pv.infn.it}
\emailAdd{marcvdh@kph.uni-mainz.de}
\abstract
{
We present a systematic study of generalized transverse-momentum dependent parton distributions (GTMDs). By taking specific limits or projections, these GTMDs 
yield various transverse-momentum dependent and generalized parton distributions, thus 
providing a unified framework to simultaneously model different 
observables. We present such simultaneous modeling by considering a light-cone wave function overlap representation of the GTMDs. We construct the different quark-quark correlation functions from the 3-quark Fock components within both the light-front constituent quark model as well as within the chiral quark-soliton model. We provide a comparison with available data and make predictions for different observables. 
}

\keywords{Phenomenological Models, QCD, Deep Inelastic Scattering}

\begin{document}
\maketitle
\flushbottom
\section{Introduction}
\label{section-1}

\begin{figure}[t!]
\begin{center}
\epsfig{file=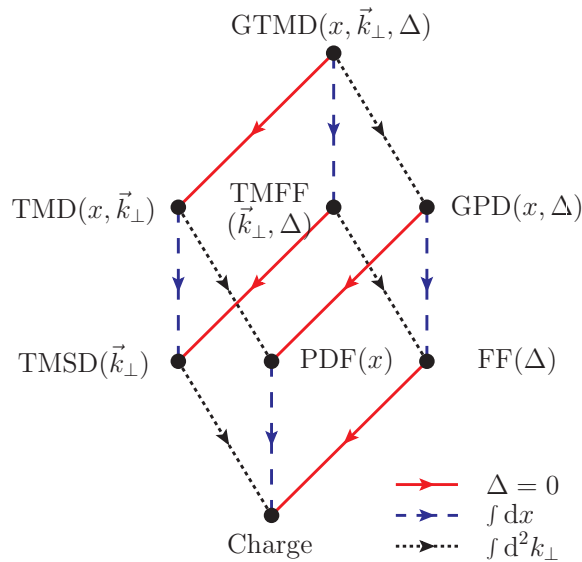,  width=0.5\columnwidth}
\end{center}
\caption{\footnotesize{Representation of the projections of the GTMDs into 
parton distributions and form factors.
The arrows correspond to different reductions in the hadron and quark momentum
space: the solid (red) arrows  give the forward limit in the hadron momentum,
the dotted (black) arrows correspond to integrating over the quark 
transverse-momentum and the dashed (blue) arrows project out the longitudinal momentum of quarks.
The different objects resulting from these links are explained in the text.}}
\label{fig11}
\end{figure}

The investigation how the composite structure of a hadron, consisting of 
near massless constituents,  results from the underlying quark-gluon dynamics,  is a 
challenging problem, as it is of non-perturbative nature, and displays many facets. 
What are the longitudinal momentum distributions of partons in a fast moving unpolarized 
or polarized hadron? What amount of transverse momentum do these partons carry, and 
how large is the resulting amount of orbital angular momentum? What is the spatial distribution 
of quarks inside a hadron as seen by a vector probe (coupling to the charge of the system), 
by an axial vector probe (coupling to the axial charge), or even seen by a more complicated probe?
\newline
\indent
The Generalized Parton Correlation Functions (GPCFs) provide a 
unified framework to address and quantify such questions. 
The GPCFs parametrize the fully unintegrated off-diagonal quark-quark correlator, depending on the full 4-momentum $k$ of the quark  and on the 4-momentum $\Delta$ which is transferred by 
the probe to the hadron; for a classification see refs.~\cite{Meissner:2008ay,Meissner:2009ww}.  
They have a direct connection with the Wigner distributions of the parton-hadron system~\cite{Ji:2003ak,Belitsky:2003nz,Belitsky:2005qn}, which represent the quantum mechanical analogues of the classical phase-space distributions.
\newline
\indent
When integrating the GPCFs over the light-cone energy component of the quark momentum one arrives at generalized transverse-momentum dependent parton distributions (GTMDs) which contain the most general one-body information of partons, corresponding to the full one-quark density matrix in momentum space. 
The GTMDs reduce to different parton distributions and form factors as is shown in figure~\ref{fig11}. 
The different arrows in this figure represent particular projections in the hadron and quark momentum space, and give the links between the matrix elements of different reduced density matrices. 
\newline
\indent
Such matrix elements can in turn be parametrized in terms of generalized parton distributions (GPDs), transverse-momentum dependent parton distributions (TMDs) and generalized form factors (FFs). 
These are the quantities which enter the description of various exclusive (GPDs), semi-inclusive (TMDs), and inclusive (PDFs) deep inelastic scattering processes, or parameterize elastic scattering processes (FFs). 
At leading twist, there are sixteen complex GTMDs, which are defined in terms of the independent polarization states of quarks and hadron. In the forward limit $\Delta=0,$ they reduce to eight TMDs which depend on the longitudinal momentum fraction $x$ and transverse momentum $\vec k_\perp$ of quarks, and therefore give access to the three-dimensional picture of the hadrons in momentum space.
On the other hand, the integration over $\vec k_\perp$ of the GTMDs leads to eight GPDs which are probability amplitudes related to the off-diagonal matrix elements of the parton density matrix in the longitudinal momentum space. After a Fourier transform of $\vec\Delta_\perp$ to the impact-parameter space, they also provide a three-dimensional picture of the hadron in a mixed momentum-coordinate space \cite{Soper:1976jc,Burkardt:2000za,Burkardt:2002hr}. The common limit of TMDs and GPDs is given by the standard parton distribution functions (PDFs), related to the diagonal matrix elements of the longitudinal-momentum density matrix for different polarization states of  quarks and hadron. The integration over $x$ leads to a bilocal operator restricted to the plane transverse to the light-cone direction and brings to the lower plane of the box in Fig~\ref{fig11}. The off-forward matrix elements of this operator can be parametrized in terms of so-called transverse-momentum dependent form factors (TMFFs). Starting from the TMFFs, we can follow the same path as in the case of the GTMDs, and at each vertex of the basis of the box of figure~\ref{fig11} we find the restricted version of the operator defining the distributions in the upper plane. Therefore, integrating out the dependence on the quark transverse momentum, we encounter matrix elements parametrized in terms of form factors (FFs), while the forward limit of TMFFs leads to transverse-momentum dependent spin densities (TMSD). Both FFs and TMSDs have the charges as common limit.

Although a variety of models has been employed to explore separately the different observables related to GTMDs, a unifying formalism for modeling the GTMDs is still missing. In order to achieve that, we will exploit the language of light-cone wave functions (LCWFs), providing a representation of nucleon GTMDs which can be easily adopted in many model calculations. In order to simplify the derivation, we will focus on the three-quark (3Q) contribution to nucleon GTMDs, postponing to future works the inclusion of higher-Fock space components. In this way, we can express  the GTMDs in a compact formula as overlap of LCWFs describing the quark content of the nucleon in the most general momentum and polarization states. Then, using the projections illustrated in figure~\ref{fig11}, we can discuss the complementary aspects encoded in the different distributions and form factors. 

The plan of the paper is as follows. In section~\ref{section-2}, we discuss the formal derivation of the LCWF overlap representation of the quark contribution to GTMDs, specializing the results to two light-cone quark models, namely the chiral quark-soliton model ($\chi$QSM) and the light-cone constituent quark model (LCCQM). In section~\ref{section-3}, we focus the discussion on the  TMDs, GPDs, PDFs, FFs and charges. In particular, we derive the general formulas obtained from the projections of GTMDs, and then we discuss and compare the predictions from both the $\chi$QSM and the LCCQM. In the last section, we draw our conclusions. Technical details and explanations about the derivation of the formulas are collected in three appendices.

\section{Formalism}\label{section-2}

\subsection{Parton Correlation Functions}

The maximum amount of information on the quark distributions inside the nucleon is contained in the fully-unintegrated quark-quark correlator $\tilde W$ for a spin-$1/2$ hadron~\cite{Ji:2003ak,Belitsky:2003nz,Belitsky:2005qn,Meissner:2009ww}, defined as
\begin{equation}\label{gencorr}
\tilde W^{[\Gamma]}_{\Lambda'\Lambda}(P,k,\Delta,N;\eta)=\frac{1}{2}\int\frac{\ud^4z}{(2\pi)^4}\,e^{ik\cdot z}\,\langle p',\Lambda'|\overline\psi(-\tfrac{z}{2})\Gamma\,\mathcal W\,\psi(\tfrac{z}{2})|p,\Lambda\rangle.
\end{equation}
This correlator is a function of the initial and final hadron light-cone helicities $\Lambda$ and $\Lambda'$, the average hadron and quark four-momenta $P=(p'+p)/2$ and $k$, respectively, and the four-momentum transfer to the hadron $\Delta=p'-p$.  In this paper, we choose to work in the symmetric light-cone frame, see figure~\ref{fig1}. The corresponding kinematics is given in appendix~\ref{LCkinematics}.
\begin{figure}[ht]
\begin{center}
\epsfig{file=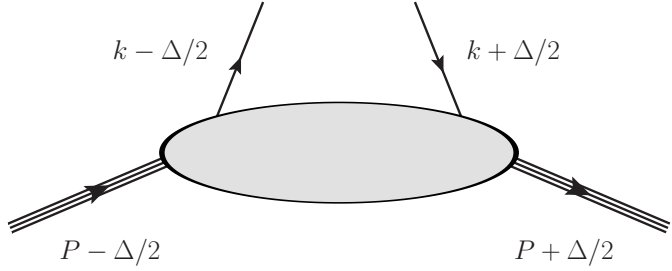,  width=0.6\columnwidth}
\end{center}
\caption{\footnotesize{Kinematics for the fully-unintegrated quark-quark correlator in a symmetric frame.}}
\label{fig1}
\end{figure} The superscript $\Gamma$ stands for any element of the basis $\{\mathds 1,\gamma_5,\gamma^\mu,\gamma^\mu\gamma_5,i\sigma^{\mu\nu}\gamma_5\}$ in Dirac space. A Wilson line $\mathcal W\equiv\mathcal W(-\tfrac{z}{2},\tfrac{z}{2}|n)$ ensures the color gauge invariance of the correlator, connecting the points $-\tfrac{z}{2}$ and $\tfrac{z}{2}$ \emph{via} the intermediary points $-\tfrac{z}{2}+\infty\cdot n$ and $\tfrac{z}{2}+\infty\cdot n$ by straight lines. This induces a dependence of the Wilson line on the light-cone direction $n$. Since any rescaled four-vector $\alpha n$ with some positive parameter $\alpha$ could be used to specify the Wilson line, the correlator actually only depends on the four-vector
\begin{equation}
N=\frac{M^2n}{P\cdot n},
\end{equation}
where $M$ is the hadron mass. The parameter $\eta=\textrm{sign}(n^0)$ gives the sign of the zeroth component of $n$, \emph{i.e.} indicates whether the Wilson line is future-pointing ($\eta=+1$) or past-pointing ($\eta=-1$).

The quark-quark correlators defining TMDs, GPDs, PDFs, FFs and charges correspond to specific limits or projections of eq.~\eqref{gencorr}. These correlators have in common the fact that the quark fields are taken at the same light-cone time $z^+=0$. Let us then focus our attention on the $k^-$-integrated version of eq.~\eqref{gencorr}
\begin{equation}\label{GTMDcorr}
\begin{split}
W^{[\Gamma]}_{\Lambda'\Lambda}&(P,x,\vec k_\perp,\Delta,N;\eta)=\int\ud k^-\,\tilde W^{[\Gamma]}_{\Lambda'\Lambda}(P,k,\Delta,N;\eta)\\
&=\frac{1}{2}\int\frac{\ud z^-\,\ud^2z_\perp}{(2\pi)^3}\,e^{ixP^+z^--i\vec k_\perp\cdot\vec z_\perp}\,\langle p',\Lambda'|\overline\psi(-\tfrac{z}{2})\Gamma\,\mathcal W\,\psi(\tfrac{z}{2})|p,\Lambda\rangle\big|_{z^+=0},
\end{split}
\end{equation}
where we used for a generic four-vector $a^\mu=[a^+,a^-,\vec a_\perp]$ the light-cone components $a^\pm=(a^0\pm a^3)/\sqrt{2}$ and the transverse components $\vec a_\perp=(a^1,a^2)$, and where $x=k^+/P^+$ and $\vec k_\perp$ are the average fraction of longitudinal momentum and average transverse momentum of the quark. A complete parametrization of this object in terms of GTMDs has been achieved in ref.~\cite{Meissner:2009ww}. GTMDs can be considered as the \emph{mother distributions} of GPDs and TMDs. Even though in the present paper we restrict our discussions to TMDs, GPDs, PDFs, FFs and charges, the formalism can readily be applied to the case of GTMDs and will be the subject of an upcoming paper.

\subsection{Overlap Representation}

Following the lines of refs.~\cite{Diehl:2000xz,Brodsky:2000xy}, we obtain in the light-cone gauge $A^+=0$ an overlap representation for the correlator~\eqref{GTMDcorr} at the twist-2 level. Moreover, this being a preliminary study, we restrict ourselves to the 3Q Fock sector\footnote{We consider here only transitions which are diagonal in both flavor and color spaces. Flavor and color indices are then omitted for clarity.} (and therefore to the region $\xi\leq x\leq 1$ with $\xi=-\Delta^+/2P^+$). Many theoretical approaches like \emph{e.g.}~\cite{Diehl:1999tr} suggest that higher Fock components yield at best a 20\% correction 
to  quark observables with up and down flavors. We then write the correlator~\eqref{GTMDcorr} as the following overlap
\begin{equation}\label{overlap}
W^{[\Gamma]}_{\Lambda'\Lambda}(P,x,\vec k_\perp,\Delta,N;\eta)=\frac{1}{\sqrt{1-\xi^2}}\sum_{\beta',\beta}\int[\ud x]_3\,[\ud^2k_\perp]_3\,\bar\delta(\tilde k)\,\psi^*_{\Lambda'\beta'}(r')\,\psi_{\Lambda\beta}(r)\,M^{[\Gamma]\beta'\beta},
\end{equation}
where the integration measures are defined as
\begin{equation}
\begin{split}
[\ud x]_3&\equiv\left[\prod_{i=1}^3\ud x_i\right]\delta\!\!\left(1-\sum_{i=1}^3x_i\right),\\
[\ud^2k_\perp]_3&\equiv\left[\prod_{i=1}^3\frac{\ud^2k_{i\perp}}{2(2\pi)^3}\right]2(2\pi)^3\,\delta^{(2)}\!\!\left(\sum_{i=1}^3\vec k_{i\perp}\right).
\end{split}
\end{equation}
The indices $\beta$ and $\beta'$ refer collectively to initial and final quark light-cone helicities $\lambda_i$ and $\lambda'_i$, respectively. The tensor $M^{[\Gamma]\beta'\beta}$ then represents the transition from the initial configuration to the final configuration of quark light-cone helicities and depends naturally on the Dirac operator $\Gamma$. The 3Q LCWF $\psi_{\Lambda\beta}(r)$ depends on $r$ which refers collectively to the momentum coordinates of the quarks in the hadron frame $\tilde k_i=(y_i,\vec\kappa_{i\perp})$, see appendix~\ref{LCkinematics}. The function $\bar\delta(\tilde k)$ selects the active quark average momentum
\begin{equation*}
\bar\delta(\tilde k)\equiv\sum_{i=1}^3\Theta(x)\,\delta(x-x_i)\,\delta^{(2)}(\vec k_\perp-\vec k_{i\perp}).
\end{equation*}

Since the labeling of quarks is arbitrary, we choose to label the active quark with $i=1$ and the spectator quarks with $i=2,3$. We can then write
\begin{equation}
\bar\delta(\tilde k)=3\,\Theta(x)\,\delta(x-x_1)\,\delta^{(2)}(\vec k_\perp-\vec k_{1\perp})
\end{equation}
and
\begin{equation}
M^{[\Gamma]\beta'\beta}=M^{[\Gamma]\lambda'_1\lambda_1}\,\delta^{\lambda'_2\lambda_2}\,\delta^{\lambda'_3\lambda_3}\qquad\text{with}\qquad M^{[\Gamma]\lambda'\lambda}\equiv\frac{\overline{u}(p',\lambda')\Gamma u(p,\lambda)}{2P^+\sqrt{1-\xi^2}},
\end{equation}
where $u(p,\lambda)$ is the free light-cone Dirac spinor. The light-cone helicity of a spectator quark is always conserved.

\subsection{Helicity and Four-Component Bases}

There are only four twist-two Dirac structures $\Gamma_\text{twist-2}=\{\gamma^+,i\sigma^{1+}\gamma_5,i\sigma^{2+}\gamma_5,\gamma^+\gamma_5\}$. They correspond to the four kinds of transition the light-cone helicity of the active quark can undergo, see \emph{e.g.}~\cite{Boffi:2002yy,Boffi:2003yj,Pasquini:2005dk}
\begin{equation}\label{correspondence}
M^{[\gamma^+]\lambda'\lambda}=\delta^{\lambda'\lambda},\quad M^{[i\sigma^{j+}\gamma_5]\lambda'\lambda}=(\sigma_j)^{\lambda'\lambda},\quad M^{[\gamma^+\gamma_5]\lambda'\lambda}=(\sigma_3)^{\lambda'\lambda}
\end{equation}
with $\sigma_i$ the three Pauli matrices. For further convenience, we associate a four-component vector\footnote{Note this is \emph{not} a Lorentz four-vector but Einstein's summation convention still applies.} to every quantity with superscript $\Gamma$
\begin{equation}
a^{[\Gamma]}\mapsto a^\nu=\left(a^0,a^1,a^2,a^3\right)\equiv\left(a^{[\gamma^+]},a^{[i\sigma^{1+}\gamma_5]},a^{[i\sigma^{2+}\gamma_5]},a^{[\gamma^+\gamma_5]}\right).
\end{equation}
With this notation, the correspondence \eqref{correspondence} takes the simple form
\begin{equation}
M^{\nu\lambda'\lambda}=(\bar\sigma^\nu)^{\lambda'\lambda},
\end{equation}
where $\bar\sigma^\nu=(\mathds{1},\vec \sigma)$. One can think of $\bar\sigma^{\nu\lambda'\lambda}\equiv(\bar\sigma^\nu)^{\lambda'\lambda}$ as the matrix of a mere change of basis, the one being labeled by the couple $\lambda\lambda'$ and the other by $\nu$
\begin{equation}
a^\nu=\sum_{\lambda'\lambda}\bar\sigma^{\nu\lambda'\lambda}\,a_{\lambda\lambda'}.
\end{equation}

In the literature one often represents correlators in the helicity basis, \emph{i.e.} in terms of helicity amplitudes
\begin{equation}
W_{\Lambda'\lambda',\Lambda\lambda}\equiv\frac{1}{2}\,W^\nu_{\Lambda'\Lambda}\sigma_{\nu\lambda\lambda'},
\end{equation}
where $\sigma_\nu=g_{\nu\rho}\,\sigma^\rho$ with $\sigma^\nu=(\mathds{1},-\vec \sigma)$. The symbols $\bar\sigma^\mu$ and $\sigma_\mu$ satisfy the relations
\begin{equation}
\frac{1}{2}\,\bar\sigma^{\nu\lambda'\lambda}\sigma_{\nu\tau\tau'}=\delta^{\lambda'}_{\tau'}\delta^\lambda_\tau,\qquad\frac{1}{2}\uTr\left[\bar\sigma^\mu\sigma_\nu\right]=\frac{1}{2}\sum_{\lambda'\lambda}\bar\sigma^{\mu\lambda'\lambda}\sigma_{\nu\lambda\lambda'}=\delta^\mu_\nu.
\end{equation}
We find however more convenient to work in the four-component basis. We then introduce tensor correlators
\begin{equation}
W^{\mu\nu}\equiv\frac{1}{2}\uTr\left[\bar\sigma^\mu W^\nu\right]=\frac{1}{2}\sum_{\Lambda'\Lambda}\bar\sigma^{\mu\Lambda\Lambda'}W^\nu_{\Lambda'\Lambda}.
\end{equation}
Helicity amplitudes and tensor correlators are related as follows
\begin{equation}
W_{\Lambda'\lambda',\Lambda\lambda}=\frac{1}{2}\,W^{\mu\nu}\sigma_{\mu\Lambda'\Lambda}\sigma_{\nu\lambda\lambda'},\qquad
W^{\mu\nu}=\frac{1}{2}\sum_{\Lambda'\Lambda\lambda'\lambda}\bar\sigma^{\mu\Lambda\Lambda'}\bar\sigma^{\nu\lambda'\lambda}W_{\Lambda'\lambda',\Lambda\lambda}.
\end{equation}

\subsection{3Q LCWF from Constituent Quark Models}

So far, the exact 3Q LCWF $\psi_{\Lambda\beta}(r)$ cannot be derived directly from the QCD Lagrangian. Nevertheless, we can try to reproduce the gross features at low scales using constituent quark models. We focus here on two phenomenologically successful models which also have the advantage of incorporating consistently relativistic effects: the light-cone constituent quark model (LCCQM)~\cite{Boffi:2002yy,Boffi:2003yj,Pasquini:2005dk} and the chiral quark-soliton model ($\chi$QSM)~\cite{Diakonov:1984vw,Diakonov:1985eg,Diakonov:1986yh,Diakonov:1987ty,Petrov:2002jr,Diakonov:2005ib,Lorce:2006nq,Lorce:2007as,Lorce:2007fa}. In the LCCQM, one describes the baryon system in terms of the overlap of the baryon state with a state made of three free on-shell valence quarks. The 3Q state is however not on-shell $M\neq\mathcal M_0=\sum_i\omega_i$, where $\omega_i$ is the energy of free quark $i$ and $M$ is the physical mass of the bound state. Since the exact baryon state is unknown, one approximates the overlap by a simple analytic function and fits the free parameters in order to reproduce at best some experimental observables, like \emph{e.g.} the anomalous magnetic moment and the axial charge. In the $\chi$QSM quarks are not free but bound by a relativistic chiral mean field (semi-classical approximation). This chiral mean field creates a discrete level in the one-quark spectrum and distorts at the same time the Dirac sea. It has been shown that the distortion can be represented by additional quark-antiquark pairs in the baryon~\cite{Petrov:2002jr}. Even though the $\chi$QSM naturally incorporates higher Fock states, we restrict the present study to the 3Q sector. The inclusion of higher Fock states is postponed to a future work.

Despite the apparent differences between the LCCQM and the $\chi$QSM, it turns out that the corresponding LCWFs are very similar in structure. In both models, the spin-flavor part of the wave function is separated from the momentum part. Moreover, canonical spin and light-cone helicity are simply connected by an $SU(2)$ rotation. We can take advantage of this similarity in structure and develop a formalism in terms of the generic 3Q LCWF (remember that $\beta=\{\lambda_i\}$)
\begin{equation}\label{LCWF}
\psi_{\Lambda\beta}(r)=\mathcal N\,\Psi(r)\sum_{\sigma_i}\Phi_\Lambda^{\sigma_1\sigma_2\sigma_3}\prod_{i=1}^3 D_{\lambda_i\sigma_i}(\tilde k_i),
\end{equation}
where $\mathcal N$ is a (real) normalization factor, $\Psi(r)$ is a symmetric momentum wave function, $\Phi_\Lambda^{\sigma_1\sigma_2\sigma_3}$ is the $SU(6)$ spin-flavor wave function, and $D(\tilde k)$ is an $SU(2)$ matrix relating light-cone helicity $\lambda$ to canonical spin $\sigma$
\begin{equation}\label{generalizedMelosh}
D(\tilde k)=\frac{1}{|\vec K|}\begin{pmatrix}K_z&K_L\\-K_R&K_z\end{pmatrix},\qquad K_{R,L}=K_x\pm iK_y.
\end{equation}
In order to specify the forthcoming expressions to one of the models, one has to perform the substitutions given in table~\ref{Modelsubstitutions}.
\begin{table}[t!]
\begin{center}
\caption{\footnotesize{Expression of the 3Q LCWF in the LCCQM and the $\chi$QSM. In the LCCQM, $m$ is the constituent quark mass, $\mathcal M_0$ is the free invariant mass of the 3Q state, and $\omega$ is the free quark energy. In the $\chi$QSM, $\kappa_\perp=|\vec\kappa_\perp|$, $\mathcal M_N$ is the soliton mass, and $E_\text{lev}$ is the energy of the one-quark discrete level. The functions $f_{/\!\!/}(y,\kappa_\perp)$ and $f_\perp(y,\kappa_\perp)$ are shown in appendix~\ref{Models}.}}\label{Modelsubstitutions}
\begin{tabular}{l@{\qquad}c@{\quad}c@{\quad}c@{\quad}c}\whline
Model&$\Psi(r)$&$K_z$&$\vec K_\perp$&$\kappa_z$\\
\hline
LCCQM&$\tilde\psi(r)$&$m+y\mathcal M_0$&$\vec\kappa_\perp$&$y\mathcal M_0-\omega$\\
$\chi$QSM&$\prod_{i=1}^3|\vec K_i|$&$f_{/\!\!/}(y,\kappa_\perp)$&$\vec\kappa_\perp\,f_\perp(y,\kappa_\perp)$&$y\mathcal M_N-E_\text{lev}$\\
\whline
\end{tabular}
\end{center}
\end{table}
In the LCCQM, the quarks are free and the matrix $D(\tilde k)$ then simply corresponds to the Melosh rotation $R_{cf}$~\cite{Melosh:1974cu}
\begin{equation}\label{MatMelosh}
D_{\lambda\sigma}(\tilde k):=D^{1/2*}_{\sigma\lambda}(R_{cf}(\tilde k))=\frac{\langle\lambda|m+y\mathcal M_0+i\vec\sigma\cdot(\hat e_z\times\vec\kappa_\perp)|\sigma\rangle}{\sqrt{(m+y\mathcal M_0)^2+\vec\kappa_\perp^2}}.
\end{equation}
In the $\chi$QSM, the quarks are bound and the matrix $D(\tilde k)$ is related to the one-quark discrete-level wave function $F_{\lambda\sigma}(\tilde k)$ created by the mean field~\cite{Diakonov:2005ib}
\begin{equation}
|\vec K|\,D_{\lambda\sigma}(\tilde k):=F_{\lambda\sigma}(\tilde k).
\end{equation}
The relation between light-cone helicity and canonical spin then involves the dynamics which is encoded in the functions $f_{/\!\!/}(y,\kappa_\perp)$ and $f_\perp(y,\kappa_\perp)$. Note that in both models the rotation is about the axis\footnote{This is not surprising by noticing that the generators of transverse boosts on the light cone are given by $\vec B^\text{LC}_\perp=\frac{1}{\sqrt{2}}\left(\vec B_\perp+\vec J_\perp\times\hat e_z\right)$, where $\vec B_\perp$ and $\vec J_\perp$ are the ordinary boost and rotation generators~\cite{Kogut:1969xa}.} $\hat\kappa_\perp\times\hat e_z$. For more details on the 3Q LCWF in the LCCQM and the $\chi$QSM, see appendix~\ref{Models}.

Connecting light-cone helicity to canonical spin (and consequently light-cone wave functions to instant-form wave functions) is usually an extremely difficult task since it involves boosts which contain the interaction. Only in simplified pictures is this connection tractable. In the two models we consider here, quarks are free or interact with a relativistic mean field, \emph{i.e.} they are quasi-independent. The main effect of boosts is creating an angle between instant-form and light-cone polarizations. This angle vanishes when the particle has no transverse momentum. An equivalent point of view is to say that a quark state with definite light-cone helicity corresponds to a linear combination of quark states with both canonical spin $\uparrow$ and $\downarrow$, responsible for the non-diagonal elements in $D(\tilde k)$.

\subsection{3Q Proton Amplitude}\label{masterformula}

Inserting the LCWF given by eq.~\eqref{LCWF} in the overlap representation of the correlator tensor $W^{\mu\nu}$~\eqref{overlap}, we obtain
\begin{equation}\label{master1}
W^{\mu\nu}(P,x,\vec k_\perp,\Delta,N;\eta)=\frac{\mathcal{N}^2}{\sqrt{1-\xi^2}}\int[\ud x]_3\,[\ud^2k_\perp]_3\,\bar\delta(\tilde k)\,\Psi^*(r')\,\Psi(r)\,\mathcal A^{\mu\nu}(r',r),
\end{equation}
where $\mathcal A^{\mu\nu}(r',r)$ stands for
\begin{equation}\label{master2}
\mathcal A^{\mu\nu}(r',r)=A\,O_1^{\mu\nu}\left(l_2\cdot l_3\right)+B\left[l_2^\mu\left(l_3\cdot O_1\right)^\nu+l_3^\mu\left(l_2\cdot O_1\right)^\nu\right].
\end{equation}
The coefficients $A$ and $B$ are flavor factors. We used $l^\mu_i=O^{\mu0}_i$ with the matrix $O^{\mu\nu}$ given by
\begin{equation}\label{spinhelicity}
O^{\mu\nu}=\frac{1}{|\vec K'||\vec K|}\begin{pmatrix}
\vec K'\cdot\vec K&i\left(\vec K'\times\vec K\right)_x&i\left(\vec K'\times\vec K\right)_y&-i\left(\vec K'\times\vec K\right)_z\\
i\left(\vec K'\times\vec K\right)_x&\vec K'\cdot\vec K-2K'_xK_x&-K'_xK_y-K'_yK_x&K'_xK_z+K'_zK_x\\
i\left(\vec K'\times\vec K\right)_y&-K'_yK_x-K'_xK_y&\vec K'\cdot\vec K-2K'_yK_y&K'_yK_z+K'_zK_y\\
i\left(\vec K'\times\vec K\right)_z&-K'_zK_x-K'_xK_z&-K'_zK_y-K'_yK_z&-\vec K'\cdot\vec K+2K'_zK_z
\end{pmatrix}.
\end{equation}
Details on the derivation can be found in appendix~\ref{spinhelrot}. 

Let us interpret this master formula. The correlator tensor $W^{\mu\nu}$ has two indices $\mu$ and $\nu$ which refer to the transition undergone by the hadron  and active quark light-cone helicities, respectively. eq.~\eqref{master1} expresses the tensor correlator in terms of the 3Q overlap $\frac{\mathcal N^2}{\sqrt{1-\xi^2}}\int[\ud x]_3\,[\ud^2k_\perp]_3$ of initial $\Psi(r)$ and final $\Psi^*(r')$ symmetric momentum wave functions with the tensor $\mathcal A^{\mu\nu}(r',r)$ for a fixed mean momentum of the active quark $\Delta(\tilde k)$. The tensor $\mathcal A^{\mu\nu}(r',r)$ corresponds to the overlap of the three initial quarks with the three final ones. In the $SU(6)$ spin-flavor wave function $\Phi_\Lambda^{\sigma_1\sigma_2\sigma_3}$ one of the quark canonical spins $\sigma_i$ is always aligned with the hadron helicity $\Lambda$. Hence $\mathcal A^{\mu\nu}$ is the sum of two contributions: the hadron helicity can be aligned with the canonical spin of either the active quark $i=1$ or one of the spectator quarks $i=2,3$. Since the spectator quarks are equivalent, they enter in a symmetric way in eq.~\eqref{master2}. The coefficients $A$ and $B$ then give the weight of each contribution and depend on the flavor of the active quark and the nature of the spin-$1/2$ hadron. In a proton, we have
\begin{equation}
A_p^u=4,\qquad B_p^u=1,\qquad A_p^d=-1,\qquad B_p^d=2.
\end{equation}
Finally, the matrix $O^{\mu\nu}$ in eq.~\eqref{spinhelicity} describes the overlap of an initial quark with a final quark. Rows and columns correspond to the type of transition undergone by the quark canonical spin and light-cone helicity, respectively.

For example, consider the vector operator $\Gamma=\gamma^+$. eq.~\eqref{correspondence} tells us that this operator is not sensitive to the active quark light-cone helicity. To see what happens in terms of active quark canonical spin, we have to look at the first column of eq.~\eqref{spinhelicity}. It appears that the vector operator is generally sensitive to the canonical spin ($O^{j0}\neq 0$ for $j=1,2,3$). Note that in absence of momentum transfer $\vec K'=\vec K$, the vector operator becomes also insensitive to the active quark canonical spin. This can be easily understood as follows. The orientation of canonical polarization relative to light-cone polarization depends on the  momentum of the particle. In presence of momentum transfer, the initial and final $SU(2)$ rotations $D(\tilde k)$ are different and the vector operator effectively sees the difference between canonical spin $\uparrow$ and $\downarrow$.

\section{Results and Discussion}\label{section-3}

We now apply and specialize the general formalism described in the previous sections to the extraction of TMDs, GPDs, PDFs, FFs and charges. The normalization constant $\mathcal N$ for each model has been fixed so as to satisfy the 
valence-quark sum rules. Namely, there are (in an effective way) two up quarks and one down quark in a proton. 
In the $\chi$QSM, the one-quark discrete level wave function is given by the sum of a bare discrete level contribution $F^\text{lev}$  and a relativistic contribution $F^\text{sea}$ due to the distortion of the Dirac sea. The latter one 
is usually not discussed because it is expected to give only a small correction.
 Here we calculate this contribution  for the first time, as explained in appendix~\ref{Models}, and we explcitly check that it is in general not relevant.
Therefore, except for the charges, all the results in the $\chi$QSM  will refer only to the $F^\text{lev}$ contribution.

\subsection{Transverse Momentum-Dependent Distributions}

The forward limit $\Delta=0$ of the correlator $W$ is given by the quark-quark correlator, denoted as $\Phi$
\begin{equation}
\begin{split}
\Phi^{[\Gamma]}_{\Lambda'\Lambda}(P,x,\vec k_\perp,N;\eta)&=W^{[\Gamma]}_{\Lambda'\Lambda}(P,x,\vec k_\perp,0,N;\eta)\\
&=\frac{1}{2}\int\frac{\ud z^-\,\ud^2z_\perp}{(2\pi)^3}\,e^{ixP^+z^--i\vec k_\perp\cdot\vec z_\perp}\,\langle P,\Lambda'|\overline\psi(-\tfrac{z}{2})\Gamma\,\mathcal W\,\psi(\tfrac{z}{2})|P,\Lambda\rangle\Big|_{z^+=0}.
\end{split}
\end{equation}
It is parametrized by TMDs at leading twist in the following way
\begin{equation}\label{TMDs}
\Phi^{\mu\nu}=\begin{pmatrix}
f_1&\frac{k_y}{M}\,h^\perp_1&-\frac{k_x}{M}\,h^\perp_1&0\\
\frac{k_y}{M}\,f^\perp_{1T}&h_1+\frac{k^2_x-k^2_y}{2M^2}\,h^\perp_{1T}&\frac{k_xk_y}{M^2}\,h^\perp_{1T}&\frac{k_x}{M}\,g_{1T}\\
-\frac{k_x}{M}\,f^\perp_{1T}&\frac{k_xk_y}{M^2}\,h^\perp_{1T}&h_1-\frac{k^2_x-k^2_y}{2M^2}\,h^\perp_{1T}&\frac{k_y}{M}\,g_{1T}\\
0&\frac{k_x}{M}\,h^\perp_{1L}&\frac{k_y}{M}\,h^\perp_{1L}&g_{1L}
\end{pmatrix}.
\end{equation}
TMDs are functions of $x$ and $\vec k_\perp^2$ only. The multipole pattern in $\vec k_\perp$ is clearly visible in eq.~\eqref{TMDs}. The TMDs $f_1$, $g_{1L}$ and $h_1$ give the strength of monopole contributions and correspond to matrix elements without a net change of helicity between the initial and final states. The TMDs $f^\perp_{1T}$, $g_{1T}$, $h^\perp_1$, and $h^\perp_{1L}$ give the strength of dipole contributions and correspond to matrix elements involving one unit of helicity flip, either on the nucleon side ($f^\perp_{1T}$ and $g_{1T}$) or on the quark side ($h^\perp_1$ and $h^\perp_{1L}$). Finally, the TMD $h^\perp_{1T}$ gives the strength of the quadrupole contribution and corresponds to matrix elements where both the nucleon and quark helicities flip, but in opposite directions.
Conservation of total angular momentum tells us that helicity flip is compensated by a change of orbital angular momentum \cite{Pasquini:2008ax} which manifests itself by powers of $\vec k_\perp/M$ with $M$ the mass of the nucleon. In the spirit of \cite{Avakian:2010br}, we define the transverse $(0)$-, $(1/2)$- and $(1)$-moments of a generic TMD $j(x,\vec k_\perp^2)$ as
\begin{equation}
\begin{split}
j^{(0)}(x)&\equiv\int\ud^2k_\perp\,j(x,\vec k_\perp^2),\\
j^{(1/2)}(x)&\equiv\int\ud^2k_\perp\,\frac{k_\perp}{M}\,j(x,\vec k_\perp^2),\\
j^{(1)}(x)&\equiv\int\ud^2k_\perp\,\frac{k_\perp^2}{2M^2}\,j(x,\vec k_\perp^2).
\end{split}
\end{equation}
The transverse $(1/2)$- and $(1)$-moments are chosen such that they directly represent the strength of the dipole and quadrupole distributions as function of $x$, respectively. Note that our definition of the $(1/2)$-moment is twice larger than in \cite{Avakian:2010br}.

In absence of momentum transfer, \emph{i.e.} $\Delta=0$, the matrix $O^{\mu\nu}$ in \eqref{spinhelicity} reduces to\footnote{This matrix is orthogonal and composed of two blocks $
O=\left(\begin{smallmatrix}
1&0\\0&R
\end{smallmatrix}\right)$ where $R$ is an $SO(3)$ matrix. This is hardly surprising since in this case the transformation $O=D^\dag(\tilde k)MD(\tilde k)$ is just the well-known homomorphism between $SU(2)$ and $SO(3)$.}
\begin{equation}\label{Ored}
O^{\mu\nu}\stackrel{\Delta=0}{=}\frac{1}{\vec K^2}\begin{pmatrix}
\vec K^2&0&0&0\\
0&\vec K^2-2K^2_x&-2K_xK_y&2K_xK_z\\
0&-2K_yK_x&\vec K^2-2K^2_y&2K_yK_z\\
0&-2K_zK_x&-2K_zK_y&-\vec K^2+2K^2_z
\end{pmatrix},
\end{equation}
and the tensor $\mathcal A^{\mu\nu}$ of eq.~\eqref{master2} becomes
\begin{equation}
\mathcal A^{00}=\left(A+2B\right),\qquad \mathcal A^{0j}=\mathcal A^{j0}=0,\qquad\mathcal A^{ij}=A\,O^{ij}.
\end{equation}
Note that $\mathcal A^{0j}=\mathcal A^{j0}=0$
because we are not considering gluon degrees of freedom, \emph{i.e.} the generic quark wave function \eqref{LCWF} leads to vanishing Sivers and Boer-Mulders functions. Moreover, the distributions for different flavors are just proportional, which is a consequence of the underlying $SU(6)$ spin-flavor symmetry. Let us then introduce the spin-flavor factors $N^q$ and $P^q$ for a flavor $q$ as follows
\begin{equation}
N^q=\left(A^q+2B^q\right)/3\qquad\text{and}\qquad P^q=A^q/3.
\end{equation}
With this definition $N^q$ and $P^q$ can be identified, respectively, with the number of quarks of flavor $q$ in the baryon and the non-relativistic contribution to the total spin of the baryon coming from quarks of flavor $q$. For a proton, we have
\begin{equation}
\label{eq:isospin}
N_p^u=2,\qquad N_p^d=1,\qquad P_p^u=4/3,\qquad P_p^d=-1/3.
\end{equation}

\begin{figure}[t!]
\begin{center}
\epsfig{file=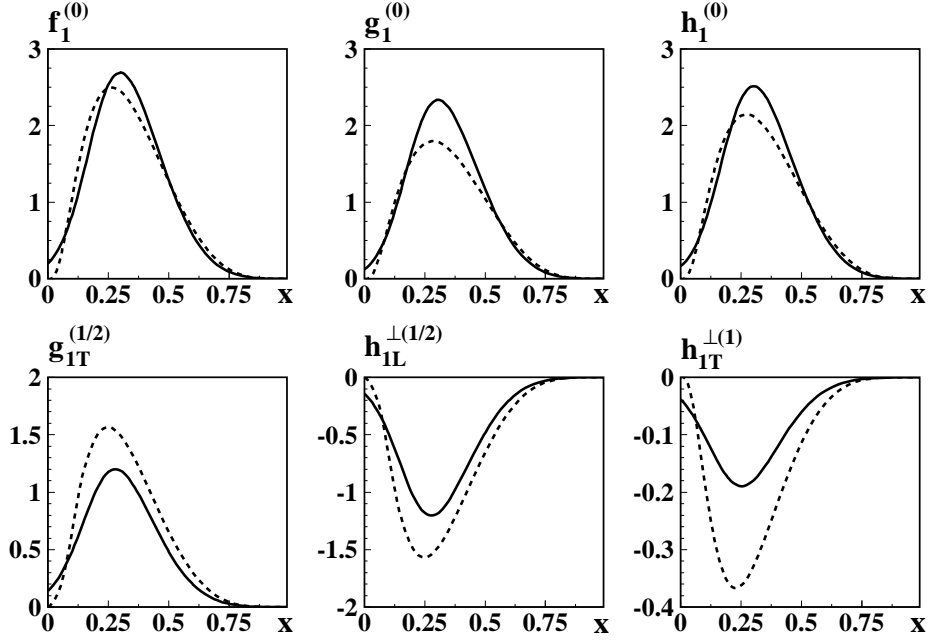,   width=0.8\columnwidth}
\end{center}
\caption{\footnotesize{Some examples of transverse moments of TMDs as function of $x$ (see text for their definition). In all panels the results are for the ``flavorless'' TMDs. TMDs of definite flavor follow from multiplication by the spin-flavor factor $N^q$ in the unpolarized case and $P^q$ in the polarized case (see eq.~\ref{eq:isospin}). Solid curves: results in the $\chi$QSM.
Dashed curves: results from the LCCQM of ref.~\cite{Pasquini:2008ax}. All the results are at the hadronic scale of the models.
}}
\label{fig2}
\end{figure}

Our model results for some transverse moments of TMDs are shown in Fig~\ref{fig2}. The shape of the curves within the $\chi$QSM and the LCCQM are very similar. The size of the longitudinal and transversity distributions is somewhat smaller for the LCCQM. On the other hand, the peak of the distributions for the transverse moments of the polarized TMDs is larger in the LCCQM than in the $\chi$QSM, especially for $h_{1T}^{\perp(1)}$. 
This pattern for the magnitude of TMDs in the two models indicates that there is more orbital angular momentum in the LCWF of the LCCQM than in  the $\chi$QSM.
This can be understood as follows. In these two models, the instant-form wave functions do not contain any orbital angular momentum. The proton spin originates solely from the quark spins. However, the corresponding light-cone wave functions do involve orbital angular momentum since quark canonical spin and light-cone helicity do not coincide in general. The proton spin originates from \emph{both} the quark light-cone helicities and orbital angular momentum. In other words, all the orbital angular momentum that appears in this paper comes from the generalized Melosh rotation of eq.~\eqref{generalizedMelosh}, or more precisely from its non-diagonal elements $\vec K_\perp$.
As one can see from eqs.~\eqref{TMDs} and \eqref{Ored}, all the functions $(g_{1L}-h_1)$, $g_{1T}$, $h^\perp_{1L}$ and $h^\perp_{1T}$ vanish identically in the limit $K_\perp\to 0$, \emph{i.e.} in absence of orbital angular momentum. It is then hardly surprising that the TMDs we obtained are not all independent. There exist three relations among polarized TMDs, which are flavor independent and read
\begin{gather}
g_{1T}+h^\perp_{1L}=0,\label{relation-pol1}\\
g_{1L}-\left[h_1+\frac{k^2_\perp}{2M^2}\,h^\perp_{1T}\right]=0,\label{relation-pol2}\\
g^2_{1T}+2h_1h^\perp_{1T}=0.\label{relation-pol3}
\end{gather}
One further flavor-dependent relation involves both polarized and unpolarized TMDs, and is given by
\begin{equation}
{\cal D}^q f_1^q+g_{1L}^q=2h_1^q,
\label{relation-unp}
\end{equation}
where ${\cal D}^q=P^q/N^q$. For example, from the relation~\eqref{relation-pol2} we see that the difference between helicity and transversity distributions $g_{1L}-h_1$ is directly connected to the pretzelosity distribution $h^\perp_{1T}$. The difference being smaller in the $\chi$QSM, it follows immediately that the pretzelosity distribution is also smaller compared to the LCCQM.

The relations in eqs.~(\ref{relation-pol1})-(\ref{relation-unp}) actually hold in a large class of quark models (see refs.~\cite{Avakian:2010br,Avakian:2008dz} and references therein). Their physical origin has been briefly discussed in \cite{Pasquini:2010pa} and will be explained in more details in a forthcoming publication. Although such quark model relations are appealing, one should keep in mind that they break down in models with gauge-field degrees of freedom and are not preserved under QCD evolution. Despite  these limitations, they can provide useful guidelines for TMD parametrizations to be further tested in experiments. An interesting result comes from recent lattice calculations~\cite{Musch:2010ka,Hagler:2009mb}, which give $h_{1L}^{\perp(1)}\approx -g_{1T}^{(1)}$, in favor of the relation~(\ref{relation-pol1}). In particular, they obtained for the average dipole deformation  (along the direction of the nucleon polarization) of the $k_\perp$ density for longitudinally polarized quarks in a transversely polarized nucleon $\langle k_x^u(g_{1T})\rangle=M\,g_{1T}^{(1)\,u}/f_1^{(0)\,u}=67(5)$ MeV, for up quarks, and $\langle k_x^d(g_{1T})\rangle=-30(5)$ MeV, for down quarks. The corresponding values for transversely polarized quarks in a longitudinally polarized nucleon are $\langle k_x^u(h_{1L}^{\perp})\rangle=M\,h_{1L}^{\perp(1)\,u}/f_1^{(0)\,u}=-60(5)$ MeV, for up quarks, and $\langle k_x^d(h_{1L}^\perp)\rangle=16(5)$ MeV, for down quarks. These results are remarkably similar to our quark model calculations: $\langle k_x^u(g_{1T})\rangle= -\langle k_x^u(h_{1L}^\perp)\rangle=54.10$ and $55.8$ MeV, and $\langle k_x^d(g_{1T})\rangle=-\langle k_x^d(h_{1L}^\perp)\rangle=-27.05$ and $-27.9$ MeV in the $\chi$QSM and the LCCQM, respectively.

Note also that, in the $\chi$QSM, the structure functions do not vanish at $x=0$ while they do vanish in the LCCQM. In the LCCQM, the functional form of 3Q LCWF is assumed on the basis of phenomenological arguments. Using the simple power-law Ansatz, the LCWF itself vanishes when any $y_i\to 0$ (see appendix~\ref{Models}). This is at variance with the wave function of the $\chi$QSM which comes from the solution of the Dirac equation describing the motion of the quarks in the solitonic pion field.
We should also mention that the $\chi$QSM was recently applied in ref.~\cite{Wakamatsu:2009fn} to calculate the unpolarized TMDs, taking into account the whole contribution from quarks and antiquarks without expansion in the different Fock components, and obtaining results for the isoscalar combination $u+d$ which corresponds to the leading contribution in the $1/N_c$ expansion.

\subsection{Generalized Parton Distributions}

When integrating the correlator $W$ over $\vec k_\perp$, one obtains the quark-quark correlator, denoted as $F$
\begin{equation}
\begin{split}
F^{[\Gamma]}_{\Lambda'\Lambda}(P,x,\Delta,N)&=\int\ud^2k_\perp\,W^{[\Gamma]}_{\Lambda'\Lambda}(P,x,\vec k_\perp,\Delta,N;\eta)\\
&=\frac{1}{2}\int\frac{\ud z^-}{2\pi}\,e^{ixP^+z^-}\,\langle p',\Lambda'|\overline\psi(-\tfrac{z}{2})\Gamma\,\mathcal W\,\psi(\tfrac{z}{2})|p,\Lambda\rangle\Big|_{z^+=z_\perp=0}.
\end{split}
\end{equation}
Note that the integration over $\vec k_\perp$ removes the dependence on $\eta$, and we are left with a Wilson line connecting directly the points $-\tfrac{z}{2}$ and $\tfrac{z}{2}$ by a straight line. Furthermore, working in the light-cone gauge $A^+=0$, the gauge link can be ignored. The correlator $F$ is parametrized by GPDs at leading twist in the following way
\begin{equation}\label{GPDs}
F^{\mu\nu}=\begin{pmatrix}
\mathcal H&i\,\frac{\Delta_y}{2M}\,\mathcal E_T&-i\,\frac{\Delta_x}{2M}\,\mathcal E_T&0\\
i\,\frac{\Delta_y}{2M}\,\mathcal E&\mathcal H_T+\frac{\Delta^2_x-\Delta^2_y}{2M^2}\,\tilde{\mathcal H}_T&\frac{\Delta_x\Delta_y}{M^2}\,\tilde{\mathcal H}_T&\frac{\Delta_x}{2M}\,\tilde{\mathcal E}\\
-i\,\frac{\Delta_x}{2M}\,\mathcal E&\frac{\Delta_x\Delta_y}{M^2}\,\tilde{\mathcal H}_T&\mathcal H_T-\frac{\Delta^2_x-\Delta^2_y}{2M^2}\,\tilde{\mathcal H}_T&\frac{\Delta_y}{2M}\,\tilde{\mathcal E}\\
0&\frac{\Delta_x}{2M}\,\tilde{\mathcal E}_T&\frac{\Delta_y}{2M}\,\tilde{\mathcal E}_T&\tilde{\mathcal H}\end{pmatrix},
\end{equation}
where we used the ``natural'' combinations of standard GPDs\footnote{The fundamental physical object is the matrix element. 
The GPDs are defined according to a specific, but not unique, parametrization of matrix elements. Therefore they
 have no \emph{a priori} simple interpretation. Combinations at $\xi=0$ appeared already in \cite{Diehl:2005jf}
.}
\begin{equation}
\begin{aligned}
\mathcal H&=\sqrt{1-\xi^2}\left(H-\frac{\xi^2}{1-\xi^2}\,E\right),&\mathcal E&=\frac{E}{\sqrt{1-\xi^2}},\\
\tilde{\mathcal H}&=\sqrt{1-\xi^2}\left(\tilde H-\frac{\xi^2}{1-\xi^2}\,\tilde E\right),&\tilde{\mathcal E}&=\frac{\xi\,\tilde E}{\sqrt{1-\xi^2}},\\
\mathcal H_T&=\sqrt{1-\xi^2}\left[\left(H_T-\frac{\xi^2}{1-\xi^2}\,E_T\right)+\frac{\frac{\vec\Delta^2_\perp}{4M^2}\,\tilde H_T+\xi\,\tilde E_T}{1-\xi^2}\right],&\mathcal E_T&=\frac{2\tilde H_T+E_T-\xi\,\tilde E_T}{\sqrt{1-\xi^2}},\\
\tilde{\mathcal H}_T&=-\frac{\tilde H_T}{2\sqrt{1-\xi^2}},&\tilde{\mathcal E}_T&=\frac{\tilde E_T-\xi\,E_T}{\sqrt{1-\xi^2}}.
\end{aligned}
\end{equation}
Comparing eq.~\eqref{GPDs} with eq.~\eqref{TMDs}, one notices a strong analogy. 
Essentially, the same multipole pattern appears, where the role of $\vec k_\perp$ in eq.~\eqref{TMDs} is played by $\vec \Delta_\perp$ in eq.~\eqref{GPDs}. Going to impact-parameter space representation does not change the multipole structure \cite{Diehl:2005jf}.
This suggests that there might be relations or dynamical connections between GPDs and TMDs, \emph{e.g.} between $E$ and $f^\perp_{1T}$ or between $2\tilde H_T+E_T$ and $h^\perp_1$ \cite{Burkardt:2003uw,Burkardt:2003je}. 
There is however no direct link since GPDs and TMDs often originate from different mother distributions \cite{Meissner:2009ww,Meissner:2007rx}, but one might still expect some correlation between signs or similar orders of magnitude from dynamical origin (there are after all deep connections with quark orbital angular momentum).

\begin{figure}[t!]
\begin{center}
\epsfig{file=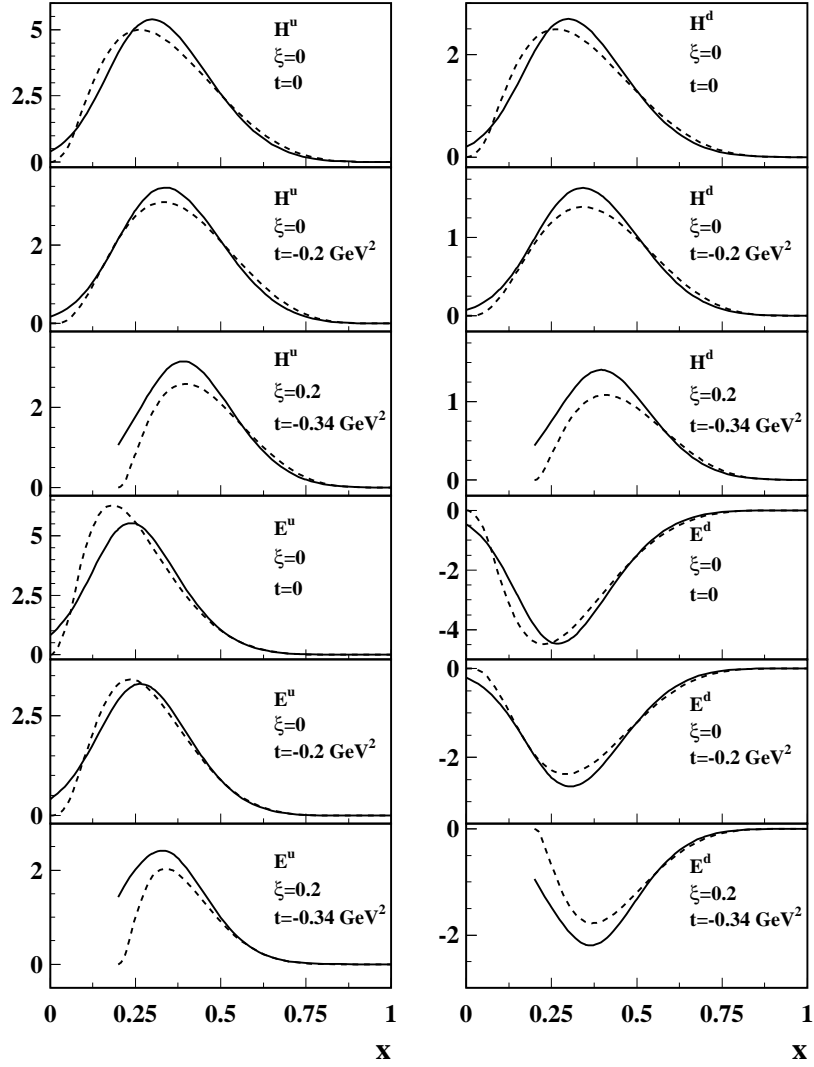,  width=0.7\columnwidth}
\end{center}
\caption{\footnotesize{Results for the 
spin averaged ($H^q$, three upper panels) and the helicity flip ($E^q$, three lower panels) generalized parton distributions for the up (left panels) and down (right panels) flavors, at fixed values of $\xi$ and $t$ as indicated.
Solid curves: results in the $\chi$QSM.
Dashed curves: results from the LCCQM of ref.~\cite {Boffi:2002yy}.}
}
\label{fig3}
\end{figure}
\begin{figure}[t!]
\begin{center}
\epsfig{file=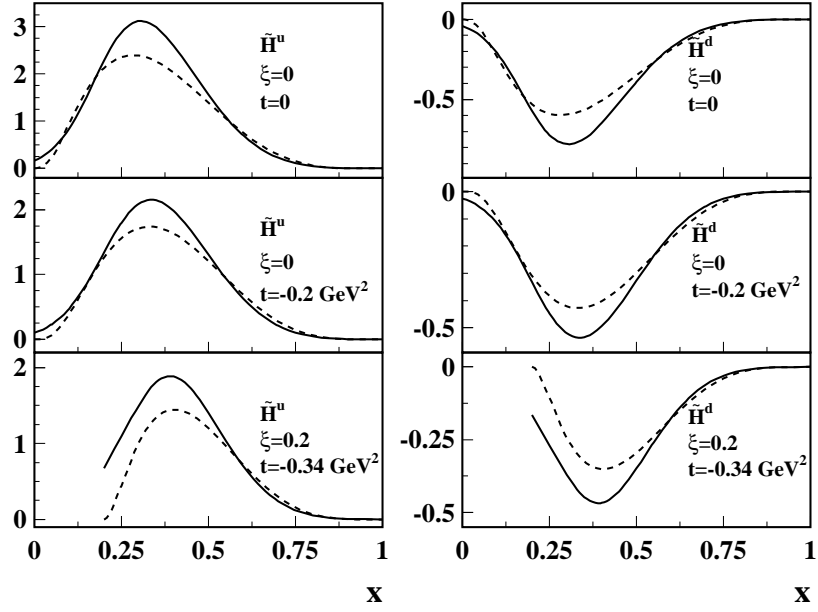,  width=0.7\columnwidth}
\end{center}
\caption{\footnotesize{Results for the helicity-dependent
generalized parton distribution $\tilde H^q$ with the same notation as in figure~\ref{fig3}. Results for the LCCQM are from ref.~\cite {Boffi:2003yj}. }
}
\label{fig4}
\end{figure}
\begin{figure}[t!]
\begin{center}
\epsfig{file=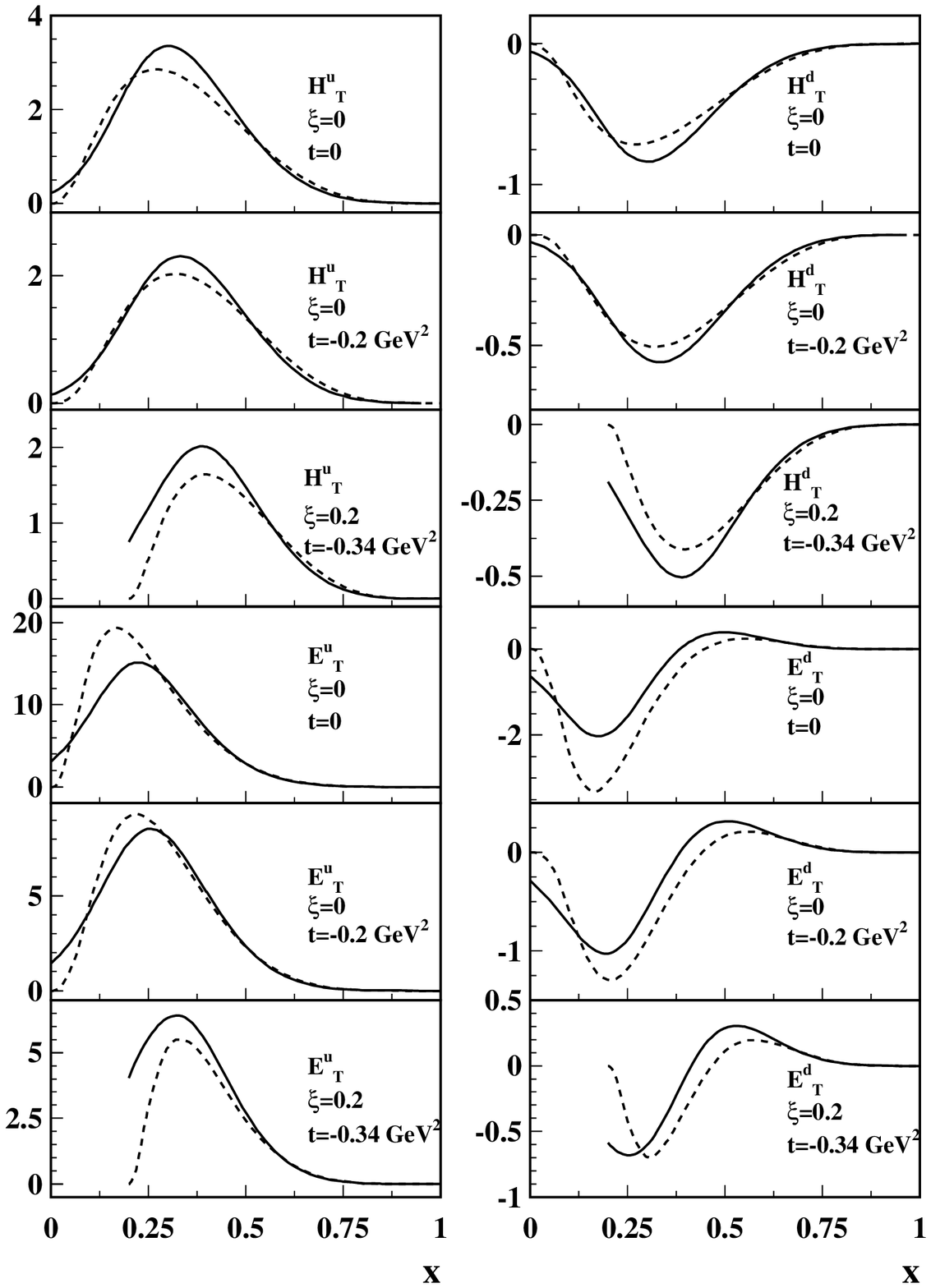,  width=0.7\columnwidth}
\end{center}
\caption{Results for the chiral-odd
$H^q_T$ (three upper panels) and $E^q_T$ (three lower panels) generalized parton distributions with the same notation as in figure~\ref{fig3}. Results for the LCCQM are from ref.~\cite {Pasquini:2005dk}.
}
\label{fig5}
\end{figure}
\begin{figure}[t!]
\begin{center}
\epsfig{file=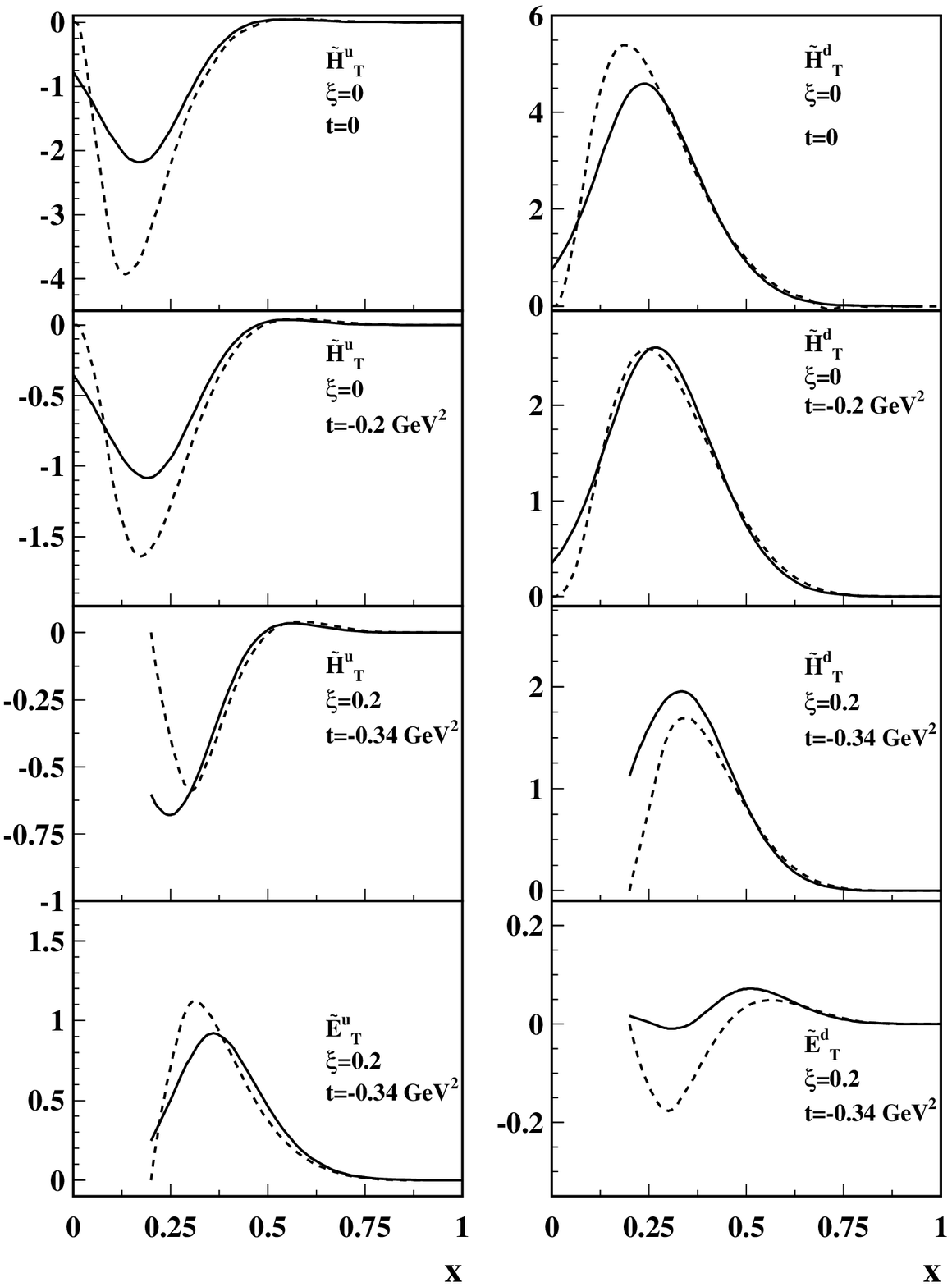,  width=0.7\columnwidth}
\end{center}
\caption{Results for the chiral-odd
$\tilde H^q_T$ (three upper panels) and $\tilde E^q_T$ (lower panels) generalized parton distributions with the same notation as in figure~\ref{fig3}. Results for the LCCQM are from ref.~\cite {Pasquini:2005dk}.}
\label{fig6}
\end{figure}

Results for the GPDs as function of $x$ and different values of $\xi$ and $t$ are shown in figures~\ref{fig3}-\ref{fig6}. Since we are considering only quark degrees of freedom, the calculations are restricted to the DGLAP region for $x\geq \xi$. The $\chi$QSM was already applied to obtain predictions for the GPDs within a different framework~\cite{Ossmann:2004bp,Polyakov:1999gs, Penttinen:1999th,Goeke:2001tz,Wakamatsu:2008ki,Wakamatsu:2006dy,Wakamatsu:2005vk}, \emph{i.e.} using the instant-form quantization and incorporating the full contribution from the discrete level and Dirac sea, without expansion in the different Fock-space components. In this way the full range of $x$ was explored, but only for the flavor combinations which correspond to the leading contribution in the $1/N_c$ expansion.

For all the GPDs we show the separate up and down-quark contributions.
Note that in contrast to TMDs, GPDs for different quark flavors are 
not simply proportional. As long as the momentum transfer $t=\Delta^2$ is not 
vanishing, the two terms in eq.~\eqref{master2} are usually different 
and non-zero. 

The point $x=\xi$ corresponds to vanishing longitudinal momentum for 
the active quark in the final state (see eq.~(\ref{eq-kin-active}) in appendix~\ref{LCkinematics}). Therefore, similarly to the case of the TMDs, 
it is a node  for the wave function in the LCCQM, but not for the $\chi$QSM. Accordingly, all the GPDs from the LCCQM are vanishing in this point.

The distributions of unpolarized quarks are described by the GPDs $H$ and $E$, the last one being non-vanishing only in the presence of orbital angular momentum.
Comparing the results from the LCCQM and the $\chi$QSM in figure~\ref{fig3}, we see that they are very similar in size and in the behavior at large $x$.
The faster fall off of $E^q$ with respect to $H^q$ for $x\rightarrow 1$ is due to the decreasing role of the orbital angular momentum for increasing longitudinal momentum of the active quark. This feature is common also to the other GPDs which arise from a transfer of orbital angular momentum between the initial and final state, like  the chiral-odd GPDs $E_T$ and $\tilde E_T$ shown in figures \ref{fig5} and \ref{fig6}, respectively. For longitudinally polarized quarks, we only show the results for the GPD $\tilde H^q$ in figure~\ref{fig4}. We refrain from presenting the quark contribution to the $\tilde E^q$ without discussing also the pion-pole term. As a matter of fact, the pion-pole is by far the largest contribution to this GPD and the differences between the LCCQM and the $\chi$QSM for the quark contribution would not be relevant for the total results. For the chiral-odd GPDs in figures~\ref{fig5} and \ref{fig6}, at $\xi=0$ there is no $\tilde E_T^q$ because it vanishes identically being an odd function of $\xi$ as consequence of time-reversal invariance.

Another common property to the model results for all GPDs concerns the $t$-dependence. In general it is more pronounced in the low $x$ region and it affects the position of the peak, especially in the LCCQM calculation where we observe  a shift to larger $x$ for increasing values of $t$. 
On the other hand, the distributions show a very weak $t$-dependence at large $x<1$, and 
are all vanishing at the end point $x=1$ as expected from momentum conservation.

\subsection{Parton Distribution Functions}\label{Evolution}

Considering the forward limit $\Delta=0$ of the correlator $F$ or, equivalently, integrating the correlator $\Phi$ over the quark transverse momentum $\vec k_\perp$ yields the quark-quark correlator $\mathcal F$
\begin{equation}
\begin{split}
\mathcal F^{[\Gamma]}_{\Lambda'\Lambda}(P,x,N)&=F^{[\Gamma]}_{\Lambda'\Lambda}(P,x,0,N)\\
&=\int\ud^2k_\perp\,\Phi^{[\Gamma]}_{\Lambda'\Lambda}(P,x,\vec k_\perp,N;\eta)\\
&=\int\ud^2k_\perp\,W^{[\Gamma]}_{\Lambda'\Lambda}(P,x,\vec k_\perp,0,N;\eta)\\
&=\frac{1}{2}\int\frac{\ud z^-}{2\pi}\,e^{ixP^+z^-}\,\langle P,\Lambda'|\overline\psi(-\tfrac{z}{2})\Gamma\,\mathcal W\,\psi(\tfrac{z}{2})|P,\Lambda\rangle\Big|_{z^+=z_\perp=0}.
\end{split}
\label{eq:correlator-f}
\end{equation}
It is parametrized by PDFs at leading twist in the following way
\begin{equation}\label{PDFs}
\mathcal F^{\mu\nu}=\begin{pmatrix}
f_1&0&0&0\\
0&h_1&0&0\\
0&0&h_1&0\\
0&0&0&g_1
\end{pmatrix}.
\end{equation}
Naturally, only the monopoles in $\vec k_\perp$ and $\vec\Delta_\perp$ survive and one obtains the well known relations between PDFs, GPDs and TMDs
\begin{equation}
\begin{split}
f_1(x)&=H(x,0,0)=\int\ud^2k_\perp\,f_1(x,\vec k_\perp^2),\\
g_1(x)&=\tilde H(x,0,0)=\int\ud^2k_\perp\,g_{1L}(x,\vec k_\perp^2),\\
h_1(x)&=H_T(x,0,0)=\int\ud^2k_\perp\,h_1(x,\vec k_\perp^2).
\end{split}
\end{equation}

\begin{figure}[h!]
\begin{center}
\epsfig{file=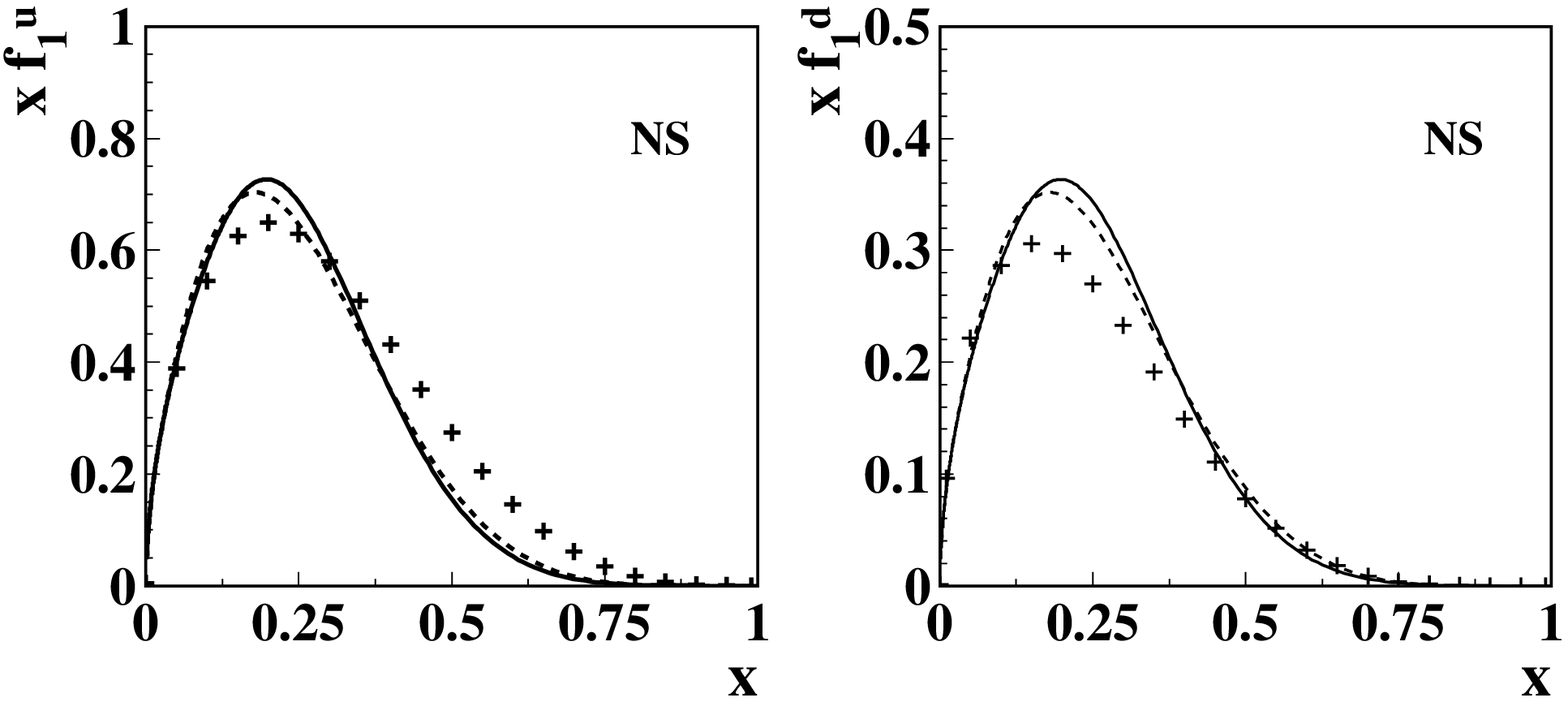,  width=.7\columnwidth}
\end{center}\caption{\footnotesize{Results for the unpolarized PDF $xf_1$ for up (left panel) and down (right panel) quark. The solid (dashed) curves are the  results from the $\chi$QSM (LCCQM) after NLO evolution from the model scale $Q^2_0=0.259$ GeV$^2$ to $Q^2=5$ GeV$^2$. The crosses are the fit to the experimental data from the CTEQ analysis at NLO of ref.~\cite{Martin:2009iq}.}}
\label{fig8}
\end{figure}

\begin{figure}[h!]
\begin{center}
\epsfig{file=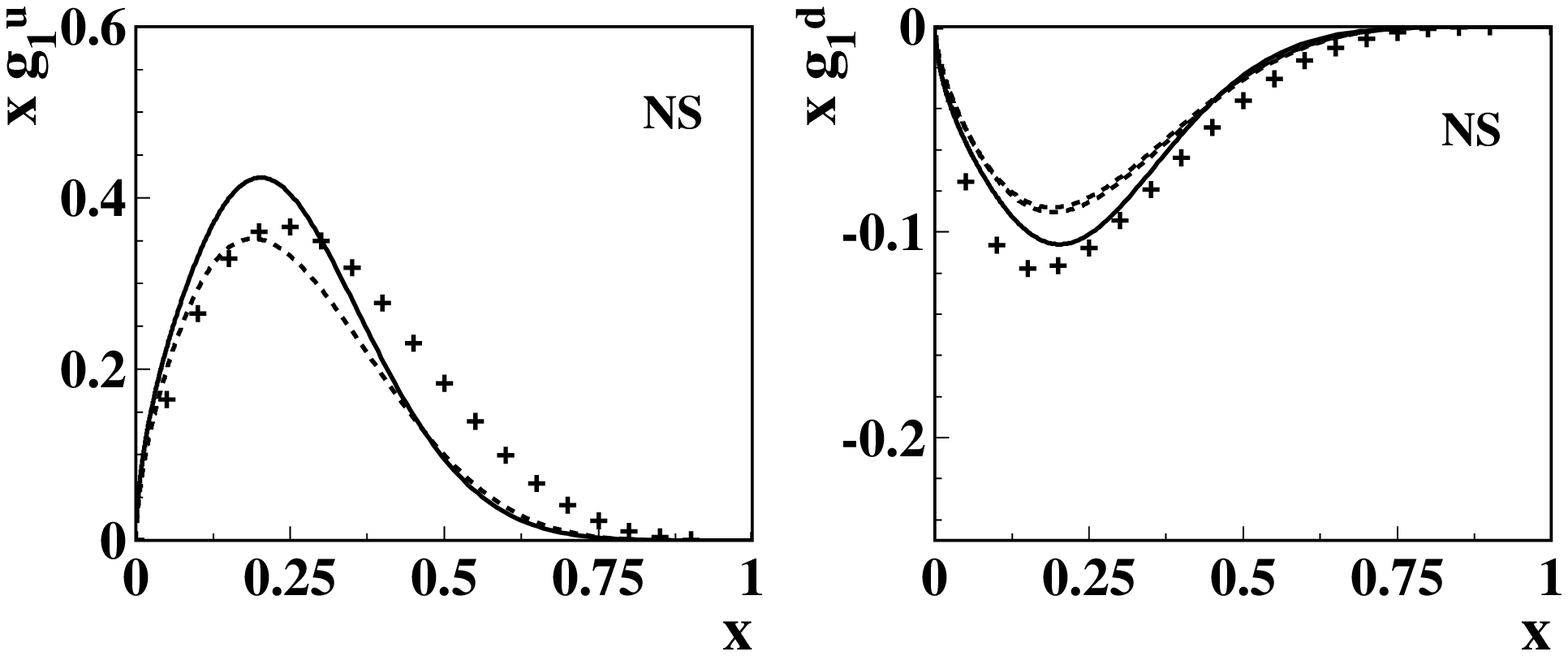,  width=.7\columnwidth}
\end{center}\caption{\footnotesize{Results for the polarized PDF $xg_1$ for up (left panel) and down (right panel) quark. The solid (dashed) curves are the  results from the $\chi$QSM (LCCQM) after NLO evolution from the model scale $Q_0^2=0.259$ GeV$^2$ to $Q^2=5$ GeV$^2$. The crosses are the fit to the experimental data from the  analysis of Ref.~\cite{Leader:2006xc}.}}\label{fig9}
\end{figure}

\begin{figure}[t!]
\begin{center}
\epsfig{file=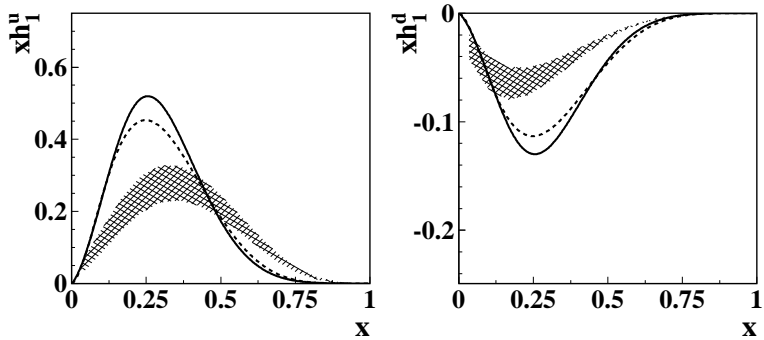,  width=.7\columnwidth}
\end{center}
\caption{\footnotesize{Results for the transversity distribution 
$xh_1$ for up (left panel) and down (right panel) quark. 
The solid (dashed) curves are the  results from the 
$\chi$QSM (LCCQM) after NLO evolution from the model scale 
$Q_0^2=0.259$ GeV$^2$ to $Q^2=2.5$ GeV$^2$. 
The shaded area corresponds the uncertainty band due to the statistical 
error of the parametrization of refs.~\cite{Anselmino:2008jk,Anselmino:2007fs}.
}}\label{fig10}
\end{figure}

In figures~\ref{fig8}-\ref{fig10} we show the results for the unpolarized, helicity and transversity distributions from the $\chi$QSM and the LCCQM after appropriate evolution from the hadronic scale of the models to the relevant experimental scales.
A key question emerging not only here but in any nonperturbative calculation concerns the scale at which the model results for the parton distributions hold. From the point of view of QCD where both quarks and gluon degrees of freedom contribute, the role of low-energy quark models is to provide initial conditions for the QCD evolution equations. Therefore, we assume the existence of a low scale $Q_0^2$ where glue and sea-quark contributions are suppressed, and the dynamics inside the nucleon is described in terms of three valence quarks confined by an effective long-range interaction. The actual value of $Q_0^2$ is fixed by evolving back the unpolarized data, until the valence distribution matches the condition that its first moment $<x>_v$ is equal to the momentum fraction carried by the valence quarks as computed in the model. Since in our models only valence quarks contribute, the matching condition is $<x(Q_0^2)>_v=1$. Starting from the initial value $<x(Q^2)>_v\approx 0.36$ at $Q^2=10$ GeV$^2$ and fixing the values of $\Lambda_{\rm{QCD}}$ and heavy-quark masses as in Ref.~\cite{Martin:2009iq}, we find $Q_0^2|_{{\rm LO}}=0.172$ GeV$^2$ and $Q_0^2|_{{\rm NLO}}=0.259$ GeV$^2$ after LO and NLO  backward evolution, respectively~\cite{Pasquini:2011tk}. In the case of the unpolarized and polarized PDFs in figures~\ref{fig8} and \ref{fig9}, respectively, we show the results only for the non-singlet (valence) contribution after NLO evolution 
to $Q^2=5$ GeV$^2$, since the models at the hadronic scale consider valence quarks, and the gluon and sea contribution at higher scales are  generated only perturbatively. The results of the $\chi$QSM and the LCCQM are very similar, for both up and down quarks, and overall reproduce the behavior of the PDFs as obtained from phenomenological parametrizations for unpolarized~\cite{Martin:2009iq} and polarized~\cite{Leader:2006xc} PDFs fitted to experimental data. However, we notice a faster falloff of the tail of the up-quark distributions at larger $x$  in our models with respect to the parametrizations. For the transversity distribution in figure~\ref{fig10}, we show the model results after NLO evolution to $Q^2=2.5$ GeV$^2$, in comparison with the available parametrization from refs.~\cite{Anselmino:2008jk,Anselmino:2007fs}. In this case, since there is no gluon counterpart and the sea can not be generated perturbatively, the model results after evolution correspond to the pure valence contribution. Compared with the phenomenological parametrization, our results are larger for both up and down quarks. 
However, we find that $h_1$ is  smaller than the Soffer bound~\cite{Soffer:1994ww} calculated from the model predictions for $f_1$ and $g_1$, \emph{i.e.} $|h^q_1(x)|\le \frac{1}{2}[f_1^q(x)+g_1^q(x)]$. The results within the LCCQM for the transversity were also used in ref.~\cite{Boffi:2009sh} to predict the Collins asymmetry in semi-inclusive deep inelastic (SIDIS) scattering. 
By using the Collins  fragmentation function $H_1^\perp$ of ref.~\cite{Efremov:2006qm}, 
the work of ref.~\cite{Boffi:2009sh} obtained a very good agreement  with  available HERMES~\cite{Diefenthaler:2005gx} and COMPASS~\cite{Alekseev:2008dn} data.
However other extractions of $H_1^\perp$  in the literature~\cite{Anselmino:2007fs,Anselmino:2008jk,Efremov:2006qm,Vogelsang:2005cs} are based on different assumptions.
In particular,
issues concerning evolution effects~\cite{Boer:2001he} in the extraction 
of the Collins function from data spanning a large range of $Q^2$
 are not yet settled.
The Collins function from~\cite{Efremov:2006qm} was extracted from a fit to 
SIDIS data~\cite{Diefenthaler:2005gx,Alekseev:2008dn}. On the other hand, the authors of Refs.~\cite{Anselmino:2007fs,Anselmino:2008jk} performed a simulatenous fit of SIDIS and  $e^+e^-$ annihilation data from BELLE~\cite{Abe:2005zx}, including approximately effects of the evolution in $Q^2$. 
 The difference in the results for $H_1^\perp$ from the two analyses explains why 
 both the results for $h_1$ in figure~\ref{fig10}, 
within the LCCQM and from the extraction~\cite{Anselmino:2007fs,Anselmino:2008jk},
 are consistent with the data.
\newline
\noindent
We also mention that  the calculation of the helicity and tansversity distribution in the $\chi$QSM, without expansion in the Fock space and using instant-forn quantization, can be found in references~\cite{Goeke:2000wv,Schweitzer:2001sr}.

\subsection{Form Factors}

\begin{figure}[t!]
\begin{center}
\epsfig{file=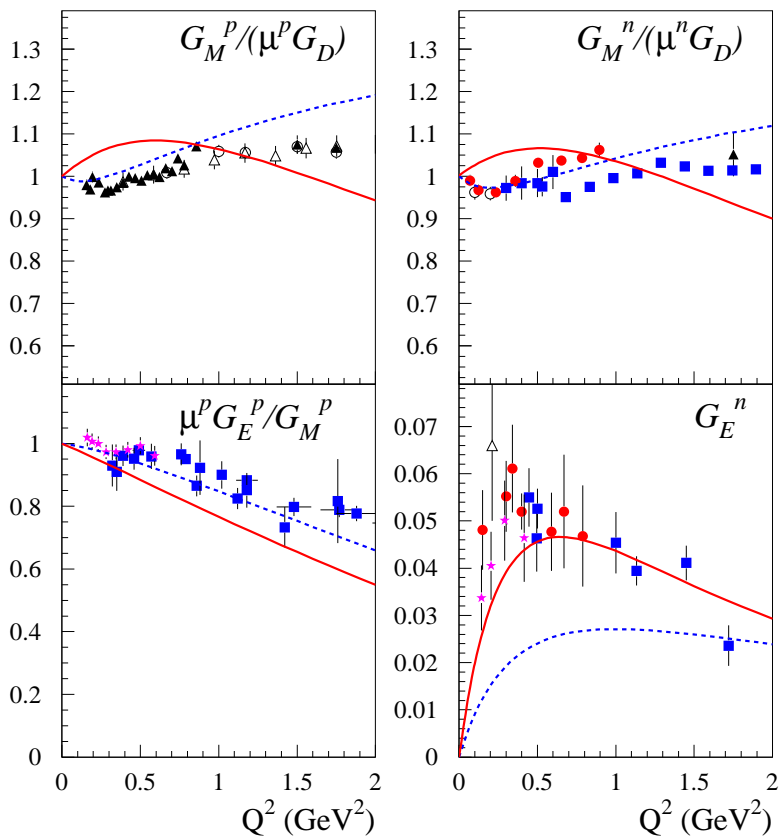,  width=0.8\columnwidth}
\end{center}
\caption{\footnotesize{The four nucleon electromagnetic Sachs FFs compared to 
the world data: MIT (purple stars), MAMI (red circles), JLab (blue squares), as well as older results (black triangles and open symbols). The references to the data can be found in \cite{Perdrisat:2006hj}.
Solid curves: results in the $\chi$QSM.
Dashed curves: results from the LCCQM of Ref.~\cite{Pasquini:2007iz}. 
The values for the anomalous magnetic moments are given in table~\ref{Chargetable}.}}
\label{fig7}
\end{figure}

When integrating the correlator $F$ over $x$, one obtains the quark-quark correlator $A$
\begin{equation}
\begin{split}
A^{[\Gamma]}_{\Lambda'\Lambda}(P,\Delta)&=\int\ud x\,F^{[\Gamma]}_{\Lambda'\Lambda}(P,x,\Delta,N)\\
&=\int\ud x\,\ud^2k_\perp\,W^{[\Gamma]}_{\Lambda'\Lambda}(P,x,\vec k_\perp,\Delta,N;\eta)\\
&=\frac{1}{2P^+}\,\langle p',\Lambda'|\overline\psi(0)\Gamma\psi(0)|p,\Lambda\rangle.
\end{split}
\end{equation}
Since in this case we deal with a local operator, the Wilson line drops out together with the dependence on the light-cone direction $n$. This correlator $A$ is parametrized by FFs in the following way
\begin{equation}
A^{\mu\nu}=\begin{pmatrix}
F_1&i\,\frac{\Delta_y}{2M}\,H_2&-i\,\frac{\Delta_x}{2M}\,H_2&0\\
i\,\frac{\Delta_y}{2M}\,F_2&H_1+\frac{\Delta^2_x-\Delta^2_y}{2M^2}\,H_3&\frac{\Delta_x\Delta_y}{M^2}\,H_3&\xi\,\frac{\Delta_x}{2M}\,G_P\\
-i\,\frac{\Delta_x}{2M}\,F_2&\frac{\Delta_x\Delta_y}{M^2}\,H_3&H_1-\frac{\Delta^2_x-\Delta^2_y}{2M^2}\,H_3&\xi\,\frac{\Delta_y}{2M}\,G_P\\
0&0&0&G_A\end{pmatrix}.
\label{correlator-a}
\end{equation}
FFs appear as the lowest $x$-moment of GPDs and are scale independent
\begin{equation}
\begin{aligned}
F_1(t)&=\int_{-1}^1\ud x\,H(x,\xi,t),&F_2(t)&=\int_{-1}^1\ud x\,E(x,\xi,t),\\
G_A(t)&=\int_{-1}^1\ud x\,\tilde H(x,\xi,t),&G_P(t)&=\int_{-1}^1\ud x\,\tilde E(x,\xi,t),\\
H_1(t)&=\int_{-1}^1\ud x\left[H_T(x,\xi,t)+\frac{\vec\Delta_\perp^2}{4M^2}\,\tilde H_T(x,\xi,t)\right],\\
H_2(t)&=\int_{-1}^1\ud x\left[2\tilde H_T(x,\xi,t)+E_T(x,\xi,t)\right],&H_3(t)&=-\int_{-1}^1\ud x\,\frac{\tilde H_T(x,\xi,t)}{2}.
\end{aligned}
\end{equation}
They are also independent of $\xi$ as a consequence of Lorentz invariance (polynomiality of GPDs). There is no FF associated to $\tilde E_T$ because time-reversal invariance implies that $\tilde E_T$ is odd in $\xi$, \emph{i.e.} $\tilde E_T(x,-\xi,t)=-\tilde E_T(x,\xi,t)$. By polynomiality, its lowest $x$-moment must then vanish\footnote{Note that the matrix element from which $G_P$ can be extracted has an explicit $\xi$ factor. By analogy, we might define formally a fourth tensor structure function as $H_4(t,\xi)=\int_{-1}^1\ud x\,\tilde E_T(x,\xi,t)/\xi$ which is not forced to vanish. It does however not correspond to any known Lorentz structure in the parametrization of the correlator $A$. In this case, the polynomiality argument does not apply \emph{a priori}, and $H_4$ might also depend on $\xi$.} \cite{Diehl:2001pm}. For illustration, the calculated nucleon electromagnetic Sachs FFs
\begin{equation}
G_E(Q^2)=F_1(Q^2)-\tau F_2(Q^2),\qquad G_M(Q^2)=F_1(Q^2)+F_2(Q^2)
\end{equation}
with $\tau\equiv Q^2/4M^2$ are compared with existing experimental data in figure~\ref{fig7}.
In particular we plot the results for $G_M/(\mu G_D)$ of the proton and neutron (upper panels), with the standard dipole form factor $G_D=1/(Q^2+\Lambda_D^2)^2$ and $\Lambda^2_D=0.71$ GeV$^2$, and the results for $\mu^p G_E^p/G_M^P$ and $G_{E}^n$ (lower panels), for the proton and neutron, respectively. The values for the magnetic moments are those of the models (see table~\ref{Chargetable} and subsequent discussion).

The LCCQM reproduces rather well the trend of the proton data  up to $Q^2\approx 1$ GeV$^2$. At higher values of the momentum transfer,  
the slope of $G_{M}^p$ is too steep and the curve deviates from the data. This effect is somehow compensated if we look at the ratio $\mu^p G_{E}^p/G_{M}^p$. Here the combined effect of a slightly overestimated $G_{M}^p$ and a slightly underestimated $G_{E}^p$ gives a result for the ratio which follows the fall-off of the data up to $Q^2\approx 2$ GeV$^2$. In the neutron case, the description of the form factors within the LCCQM is less satisfactory.
In particular $G_{E}^n$ is largely underestimated and we are not able to reproduce the slope at low $Q^2$. The neutron radius originates as a partial cancellation between the Dirac radius and the contribution of the anomalous magnetic moment (the so-called Foldy term) which dominates. In the LCCQM, the Dirac radius is much larger as compared with the $\chi$QSM, leading to a larger cancellation in the neutron radius, which reaches only $2/3$ of the value in the $\chi$QSM. 
As it was shown in Ref.~\cite{Pasquini:2007iz}, the description in the LCCQM can be
 improved by taking into account the contribution of the meson cloud of the
nucleon and by relaxing the approximation of $SU(6)$ symmetry in the model.
The combined effects of a small percentage of mixed-symmetric terms in the
nucleon LCWF and of the pion-cloud,  which mainly acts at very small
$Q^2$, provides a much improved understanding of $G_{E}^n$ within that model. 
Those effects are much less significant on the other nucleon form factors.

The results within the light-cone $\chi$QSM exhibit a peculiar behavior for the magnetic form factors.
Both in the proton and neutron case the results rise faster than the dipole form at low $Q^2$, while have a steeper fall off at higher $Q^2$.
The deviation is within $10\%$, but it goes in the opposite directions in the two ranges of $0< Q^2<1$ GeV$^2$ and  $1< Q^2<2$ GeV$^2$, in such a way that the global 
curvature is quite different from the data. On the other hand, the results for $G_{E}^n$ are in much better agreement with the data. The slope at low $Q^2$ is much steeper than in the LCCQM, going in the direction of the experimental data which support even a steeper 
rise. Finally, the results for $\mu^p G_{E}^p/G_{M}^p$ have a too fast falloff with $Q^2$, and 
deviate from the experimental data by $\approx 20\%$ in the whole range of $Q^2$.

\subsection{Charges}

Considering the forward limit $\Delta=0$ of the correlator $A$ or, equivalently, integrating the correlator $\mathcal F$ over $x$ yields the quark-quark correlator $\mathcal Q$
\begin{equation}
\begin{split}
\mathcal Q^{[\Gamma]}_{\Lambda'\Lambda}(P)&=A^{[\Gamma]}_{\Lambda'\Lambda}(P,0)\\
&=\int\ud x\,\mathcal F^{[\Gamma]}_{\Lambda'\Lambda}(P,x,N)\\
&=\int\ud x\,F^{[\Gamma]}_{\Lambda'\Lambda}(P,x,0,N)\\
&=\int\ud x\,\ud^2k_\perp\,\Phi^{[\Gamma]}_{\Lambda'\Lambda}(P,x,\vec k_\perp,N;\eta)\\
&=\int\ud x\,\ud^2k_\perp\,W^{[\Gamma]}_{\Lambda'\Lambda}(P,x,\vec k_\perp,0,N;\eta)\\
&=\frac{1}{2P^+}\,\langle P,\Lambda'|\overline\psi(0)\Gamma\psi(0)|P,\Lambda\rangle,
\end{split}
\end{equation}
which is parametrized by the charges in the following way
\begin{equation}\label{charges}
\mathcal Q^{\mu\nu}=\begin{pmatrix}
q&0&0&0\\
0&\delta q&0&0\\
0&0&\delta q&0\\
0&0&0&\Delta q
\end{pmatrix}.
\end{equation}
\begin{table}[t!]
\begin{center}
\caption{\footnotesize{
Results for the axial charge $\Delta q$, tensor charge $\delta q$, anomalous magnetic moment $\kappa^q$,  and tensor anomalous magnetic moment $\kappa_T^q$ in the $\chi$QSM and the LCCQM, compared with experimental and phenomenological values.
$\chi$QSM I refers to the calculation with $F^\text{lev}$ only, while the results labeled $\chi$QSM II and III include the effects of the Dirac sea on the discrete level, without and with Pauli-Villars regularization, respectively (see appendix~\ref{Models}). The 
results within the LCCQM are from refs.~\cite{Pasquini:2007iz,Pasquini:2007xz,Pasquini:2006iv}. 
The experimental data for the axial charges at $Q^2=5$ GeV$^2$ are from~\cite{Airapetian:2007mh} and the anomalous magnetic moments are from~\cite{Nakamura:2010zzi}. 
The model results for the tensor charges are evolved at LO to $Q^2=0.8$ GeV$^2$ for comparison with the values from 
the phenomenological extraction of Ref.~\cite{Anselmino:2008jk} .}}\label{Chargetable}
\begin{tabular}{l@{\quad}c@{\quad}c@{\quad}c@{\quad}c@{\quad}c@{\quad}c@{\quad}c@{\quad}c}\whline
Model&$\Delta u$&$\Delta d$&$\delta u$&$\delta d$&$\kappa^u$&$\kappa^d$&$\kappa_{T}^u$&$\kappa_{T}^d$\\
\hline
LCCQM           &$0.995$  &$-0.249$  &$0.90$                  &$-0.23$                  &$1.867$  &$-1.579$  &$3.98$  &$2.60$\\
$\chi$QSM I     &$1.148$  &$-0.287$  &$0.96$                  &$-0.24$                  &$1.766$  &$-1.551$  &$3.83$  &$2.58$\\
$\chi$QSM II    &$1.055$  &$-0.264$  &$0.93$                  &$-0.23$                  &$2.072$  &$-1.785$  &$4.43$  &$2.94$\\
$\chi$QSM III   &$1.118$  &$-0.279$  &$0.95$                  &$-0.24$                  &$1.902$  &$-1.647$  &$4.09$  &$2.74$\\
Exp. Value      &$0.825$  &$-0.444$  &$0.54^{+0.09}_{-0.22}$  &$-0.23^{+0.09}_{-0.16}$  &$1.673$  &$-2.033$  &--      &--\\
\whline
\end{tabular}
\end{center}
\end{table}
The vector charge $q=F_1^q(0)=\int\ud x\,f_1^q(x)$ represents the (effective) number of quarks with flavor $q$. 
The normalization constant $\mathcal N$ of the models has been chosen such that $u=2$ and $d=1$ for the proton. The axial charge 
$\Delta q=G_A^q(0)=\int\ud x\,g_1^q(x)$ represents the fraction of baryon helicity carried by the spin of quarks with flavor $q$. Finally, the tensor charge $\delta q=H_1^q(0)=\int\ud x\,h_1^q(x)$ represents the fraction of baryon transversity due to the spin of quarks with flavor $q$. From eq.~\eqref{Ored}, one can see that the axial and tensor charges are usually different. Only in the non-relativistic limit, \emph{i.e.} when there is no appreciable orbital angular momentum $\kappa_\perp\approx 0$ and thus no distinction between canonical spin and light-cone helicity, they do coincide. This is the origin of the claim that the difference between axial and tensor charges is a measure of quark orbital angular momentum. One should however keep in mind that this is not completely true in general because of higher Fock components. The FFs $F_2(t)$, $H_2(t)$, $H_3(t)$ and $G_P(t)$ being associated with dipole or quadrupole structures in $\vec\Delta_\perp$ do not have corresponding charges. Nevertheless, one can still define their limit for vanishing momentum transfer $t=0$ (as long as it is finite). For example, the anomalous magnetic moments and anomalous tensor magnetic moments are defined by $\kappa^q=F_2^q(0)$ and $\kappa_T^q=H_2^q(0)$, respectively. 

In table~\ref{Chargetable}, we compare the calculated charges and anomalous moments for a proton with the corresponding experimental values and phenomenological extractions. For the $\chi$QSM we quote the results with the bare discrete-level contribution only ($\chi$QSM I) and with the relativistic contribution coming the distortion of the Dirac sea, without ($\chi$QSM II) and with Pauli-Villars regularization ($\chi$QSM III), see appendix~\ref{Models} for more details. The corrections due to the effects of the sea are in general small, especially after regularization. Therefore, in the following discussion of the results we will refer to the values of $\chi$QSM I only.

Since the isovector combination of the axial charge $g_A=\Delta u-\Delta d$ is scale independent (as a consequence of current conservation) and since the isoscalar combination $\Delta u+\Delta d$ is weakly scale dependent, we did not evolve the model results for the axial charges. The last ones and the anomalous magnetic moments tend to overestimate the contribution from up quark and, vice-versa, to underestimate the down quark contribution. The deviations are such that they compensate in the results for the isovector axial charge $g_A$. We find $g_A=1.24$ and $g_A=1.44$ for the LCCQM and the $\chi$QSM, respectively, 
to be compared with the experimental value $g_A=1.269$ 
obtained from the HERMES data~\cite{Airapetian:2007mh}, in agreement with the value obtained from $\beta$-decay measurements~\cite{Nakamura:2010zzi}. On the other hand, for the isoscalar contribution $\Delta u+\Delta d$, the deviations go in the opposite direction, such that about $75\%$ ($85\%$) of the nucleon helicity is carried by the quark helicities in the LCCQM ($\chi$QSM). This has to be contrasted with the much smaller experimental result of about $37 \%$ at a scale around $10$~GeV$^2$~\cite{deFlorian:2009vb}. 

For the anomalous magnetic moments, the deviation is bigger for down quarks than for up quarks. This mainly affects the neutron results, for which the discrepancy from the experimental data is bigger than in the proton case. In particular, we find $\mu^p= 2.78$ and $\mu^n=-1.69$, within the LCCQM, and $\mu^p= 2.69$ and $\mu^n=-1.62$, within the $\chi$QSM, to be compared with the experimental values $\mu^p= 2.793$ and $\mu^n=-1.913$. A significant improvement is obtained for $\mu^n$ in the LCCQM by taking into account the meson-cloud contribution as in Ref.~\cite{Pasquini:2007iz}. This is at the price of slightly worse results for $g_A$ and $\mu^p$. As a matter of fact, it is in general a challenging task in light-cone quark models to reproduce simultaneously all the three quantities, see \emph{e.g.} refs.~\cite{Dziembowski:1987zp,Chung:1991st,Cardarelli:1995dc,Ma:2002ir}. While it has already been shown in refs.~\cite{Diakonov:2005ib,Lorce:2006nq,Lorce:2007as} that higher Fock components within the $\chi$QSM contribute significantly to the axial charges, it remains to be checked explicitly that the same holds for the magnetic moments. 

For the tensor charges, we evolved the model results at LO to $Q^2=0.8$ GeV$^2$ for comparison with the corresponding results of the phenomenological extraction of Ref.~\cite{Anselmino:2008jk}. In this case, both the  LCCQM and the $\chi$QSM agree very well with the phenomenological value for the down quark, while the up-quark contribution is larger by $\approx 40\%$. We note however that the values of the quark tensor charges are strongly scale dependent and may depend crucially on the choice of the initial scale of the models~\cite{Wakamatsu:2008ki}. A safer quantity to compare with is the ratio $\delta d/\delta u$ which is scale independent. In our models, the assumption of $SU(6)$ symmetry implies $\delta d/\delta u=-1/4$, which is compatible, within error bars, with $\delta d/\delta u=-0.42^{+0.0003}_{-0.20}$ from the fit of~\cite{Anselmino:2008jk}. We note that the deviation of the experimental data from the $SU(6)$ limit of $1/4$ for the ratio $\Delta d/\Delta u$ is much more significant. The experimental data at $Q^2=5$ GeV$^2$ give $\Delta d/\Delta u=-0.54$, and this value is weakly scale dependent. Therefore we expect that the inclusion of $SU(6)$ symmetry breaking terms can play an important role for the description of the longitudinal spin degrees of freedom, while it is less relevant for observables related to the transverse spin. Note that in the $\chi$QSM the $SU(6)$ symmetry holds only at the 3Q level. Let us mention that the instant-form version of the $\chi$QSM of Ref.~\cite{Ledwig:2010tu}, which includes in principle all Fock components, gives the ratios $\delta d/\delta u=-0.3$ and $\Delta d/\Delta u=-0.45$ at the model scale $Q_0^2=0.36$ GeV$^2$.

Finally, the results for the tensor anomalous magnetic moments are very similar in the LCCQM and the $\chi$QSM. The values of $\kappa_T^u/2M$ and  $\kappa_T^d/M$ are a measurement of the average distortion of the density describing the distribution in impact-parameter space of transversely polarized quarks in unpolarized nucleons.
Therefore, although the model results give $\kappa_T^u>\kappa_T^d$, the distortion in the spin density is larger for down quarks than for up quarks by about $30\%$.
The corresponding quantity in the transverse-momentum space is given by the average dipole distortion induced by the Boer-Mulders function. In this case, using  the recent LCCQM calculation of Ref.~\cite{Pasquini:2010af}, we find the same results for the relative size of the distortion in the down-quark distribution with respect to the up-quark distribution. The same arguments apply for the dipole distortions in the distributions of unpolarized quarks in transversely
 polarized nucleons as seen in impact-parameter space and momentum space. In this case the correspondence is between the average distortion measured through the 
anomalous magnetic moment $\kappa$ and the Sivers function. For $\kappa_T^q$ there exist also lattice calculations~\cite{Gockeler:2006zu} which refer to a renormalization scale $\mu^2=4$ GeV$^2$, and give the results $\kappa_T^u=3.0$ and $\kappa_T^d=1.9$. The instant-form version of the $\chi$QSM of Ref.~\cite{Ledwig:2010zq} give $\kappa_T^u=3.56$ and $\kappa_T^d=1.83$. Considering the ratio of up to down contribution, which is renormalization scale independent, we find a very good agreement between the lattice calculations ($1.51$) and our model results ($1.53$ in the LCCQM and $1.48$ in the $\chi$QSM).

\section{Conclusions}

In this work we presented a first study of GTMDs, which are quark-quark correlators where the quark fields are taken at the same light-cone time. By taking specific limits or projections 
of these GTMDs, they yield PDFs, TMDs, GPDs, FFs, and charges, 
accessible in various inclusive, semi-inclusive, exclusive, and elastic scattering processes. 
The GTMDs therefore provide a unified framework to simultaneously model these different 
observables. 
\newline
\indent
We took a first step in this modeling, by considering a light-cone wave function (LCWF) overlap 
representation of the GTMDs and by restricting ourselves to the 3Q Fock components in the 
nucleon LCWF. At twist-two level, we studied the most general 
transition which the active quark light-cone helicity can undergo in a polarized nucleon, corresponding to the general helicity amplitudes of the quark-nucleon system. We develop a formalism which is quite general and can be applied to many quark models as long as the nucleon state can be represented in terms of 3Q 
 without mutual interactions.
For the radial wave function of the quark in the nucleon, we studied two phenomenological 
successful models which include relativistic effects~: the light-cone constituent quark model (LCCQM) and the chiral quark-soliton model ($\chi$QSM). In the LCCQM, the 
nucleon LCWF was approximated by its leading Fock component, consisting of free on-shell 
valence quarks. In the $\chi$QSM, the quarks are not free but bound by a relativistic chiral 
mean field, which creates on the one hand a shift of the discrete level in the one-quark spectrum, and on the other hand also a distortion of the Dirac sea. The latter can be interpreted as arising 
due to the presence of additional quark-antiquark pairs in the nucleon LCWF. 
Therefore, the $\chi$QSM has the potential to systematically go beyond a description in 
terms of the leading 3Q Fock component in the nucleon LCWF.  
Restricting ourselves to the 3Q component in the present work, the quarks in both models being either free or interacting with a relativistic mean field, allowed us to 
connect light-cone helicity to canonical spin, and thus relate LCWFs with equal-time wave functions. It was shown that the resulting boosts, which connect both, in general introduce an angle between the light-cone helicity and the canonical spin for a quark with non-zero transverse momentum.
\newline
\indent
We firstly obtained the various TMDs in the forward limit, involving either no helicity flip, one unit of helicity flip of either quark or nucleon, or transitions where both quark and nucleon helicities flip. As the helicity flip processes are accompanied by a change of orbital angular momentum, 
the TMDs reveal such orbital angular momentum in the nucleon LCWF. We provided 
predictions for the various TMDs in both models, within the 3Q framework, 
and found several relations between TMDs for both models. The amount of orbital angular momentum in the LCCQM in general was found to be larger than in the 
$\chi$QSM. In particular, the TMD where both quark and nucleon helicities flip (pretzelosity) 
was found to be about twice as large in the LCCQM as compared with the $\chi$QSM.  
\newline
\indent
We next applied our formalism to describe the GPDs entering hard exclusive processes. 
We found a strong analogy in the multipole patterns in TMDs and GPDs, where the role of the 
quark transverse momentum $\vec k_\perp$ in TMDs is played by the 
momentum transfer to the hadron $\vec \Delta_\perp$ in the case of GPDs. 
We found a qualitative difference between both models for PDFs or GPDs  
at $x = 0$ (for $\xi = 0$), or in general for GPDs at $x = \xi$, corresponding
to
zero longitudinal momentum for the active quark in the final state. 
In the LCCQM all distributions vanish at this point, in contrast to the $\chi$QSM. 
We also compared PDFs in both models, and found that the non-singlet (valence) components of unpolarized (polarized) PDFs $f_1$ ($g_1$)  compare reasonably with the phenomenological extractions after NLO evolution to 5 GeV$^2$, but systematically undershoot the large-$x$ tail for the $u$-quark distribution.  For the transversity distribution, our results were found to be 
larger than the available parameterizations for both up and down quarks. 
\newline
\indent
We also compared the results obtained in both models for form factors and charges. 
For the electromagnetic FFs of the proton, and the neutron magnetic FF, the LCCQM 
was found to reproduce relatively well the $Q^2$ dependence up to about $Q^2 \sim 1$~GeV$^2$, whereas the $\chi$QSM shows too small magnetic radii. 
For the neutron electric FF,
the $\chi$QSM gives here a description in good agreement with the data, whereas 
 the LCCQM falls short of the data, by about a factor 2 in the range up to $Q^2 \sim 1$~GeV$^2$. This behaviour can be improved by considering 
$SU(6)$ breaking terms or higher Fock components, describing the physics of the pion cloud. 
 Finally we found for the isovector axial charges as values $g_A = 1.24$ ($1.44$) within the 
 LCCQM ($\chi$QSM) respectively. Furthermore, the quark helicities carry 
 75 \% (85 \%) of the nucleon helicity in the LCCQM ($\chi$QSM) respectively. 
For the tensor charges, both  LCCQM and the $\chi$QSM results evolved at LO to $Q^2=0.8$ GeV$^2$ agree well for the down quark with the available phenomenological extraction, 
whereas the model result for the up quark is larger by $\approx 40\%$ compared to the 
phenomenological value. 
\newline
\indent
As discussed above, the presented LCWF overlap framework for GTMDs 
can be systematically extended beyond the 3Q LCWF Fock component. Such 
an extension to the 5-quark Fock component and its application on the different observables 
discussed in this work, is an interesting subject for future work.

\section*{Acknowledgments}
We are grateful to P. Schweitzer for instructive discussions.
This work was supported in part  by
the Research Infrastructure Integrating Activity
``Study of Strongly Interacting Matter'' (acronym HadronPhysics2, Grant
Agreement n. 227431) under the Seventh Framework Programme of the
European Community, by the Italian MIUR through the PRIN 
2008EKLACK ``Structure of the nucleon: transverse momentum, transverse 
spin and orbital angular momentum''.

\appendix

\section{Light-Cone Kinematics}\label{LCkinematics}

In this paper, we work in the symmetric  infinite momentum frame, where $P^+$ is large, $\vec P_\perp=\vec 0_\perp$ and $\Delta\cdot P=0$. The four-momenta involved are then
\begin{equation}
\begin{split}
P&=\left[P^+,P^-,\vec 0_\perp\right],\\
k&=\left[xP^+,k^-,\vec k_\perp\right],\\
\Delta&=\left[-2\xi P^+,2\xi P^-,\vec\Delta_\perp\right],\\
n&=\left[0,\pm 1,\vec 0_\perp\right],
\end{split}
\end{equation}
with $P^-=\frac{M^2+\Delta_\perp^2/4}{2(1-\xi^2)P^+}$. Note that the form used for $n$ is not the most general one but leads to an appropriate definition of TMDs for semi-inclusive deep inelastic scattering and Drell-Yan processes.

For the active quark ($i=1$) the momentum coordinates in initial and final hadron frames are then given by
\begin{equation}
\label{eq-kin-active}
\begin{split}
\tilde k_1=(y_1,\kappa_{1\perp})&=\left(\frac{x_1+\xi}{1+\xi},\vec k_{1\perp}-\frac{1-x_1}{1+\xi}\,\frac{\vec\Delta_\perp}{2}\right),\\
\tilde k'_1=(y'_1,\kappa'_{1\perp})&=\left(\frac{x_1-\xi}{1-\xi},\vec k_{1\perp}+\frac{1-x_1}{1-\xi}\,\frac{\vec\Delta_\perp}{2}\right),
\end{split}
\end{equation}
while for the spectator quarks ($i=2,3$) they are given by
\begin{equation}
\label{eq-kin-spect}
\begin{split}
\tilde k_i=(y_i,\kappa_{i\perp})&=\left(\frac{x_i}{1+\xi},\vec k_{i\perp}+\frac{x_i}{1+\xi}\,\frac{\vec\Delta_\perp}{2}\right),\\
\tilde k'_i=(y'_i,\kappa'_{i\perp})&=\left(\frac{x_i}{1-\xi},\vec k_{i\perp}-\frac{x_i}{1-\xi}\,\frac{\vec\Delta_\perp}{2}\right).
\end{split}
\end{equation}
In Eqs.~\eqref{eq-kin-active} and \eqref{eq-kin-spect}, $\vec k_{i\perp}$ is the average quark transverse momentum  and $x_i$ the
average quark longitudinal momentum fraction in the symmetric frame.
\section{LCCQM and $\chi$QSM}\label{Models}

We collect in this appendix all the necessary material for the model evaluations. The normalizations of the wave functions are fixed by requiring that there are two up and one down quarks in the proton.

\subsection{LCCQM}

In the numerical evaluations within the LCCQM, we adopted a power-law form for the momentum wave function $\tilde\psi(r)$~\cite{Schlumpf:1992ce}
\begin{equation}
\tilde\psi(r)=2(2\pi)^3\sqrt{\frac{\omega_1\omega_2\omega_3}{y_1y_2y_3\mathcal M_0}}\,\frac{1}{(\mathcal M_0^2+\beta^2)^\gamma},
\label{eq:B1}
\end{equation}
with the parameters $\beta=0.607$ GeV and $\gamma=3.5$, and the quark mass $m=0.263$ GeV. These parameters have been fitted to reproduce at best the proton magnetic moment and the axial charge.
\begin{figure}[h!]
\begin{center}
\epsfig{file=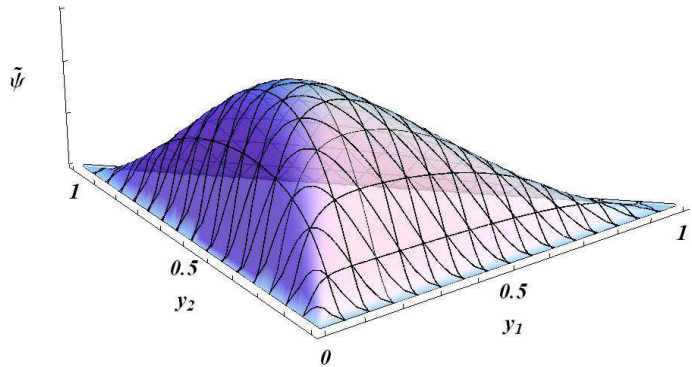, width=0.6\columnwidth}
\end{center}
\caption{\footnotesize{Power-law 3Q LCWF $\tilde\psi(r)$ used in the LCCQM as function of $y_1$ and $y_2$ with $\kappa_{1\perp}=0.2$ GeV and $\kappa_{2\perp}=0.4$ GeV.}}
\label{fig12}
\end{figure}
Since $\gamma>2$, this wave function clearly vanishes as any $y_i\to 0$, see figure~\ref{fig12}. 
In eq.~(\ref{eq:B1}) and in figure~\ref{fig12} the normalization factor is left out.
The intrinsic scale of the LCCQM is considered to be about $Q_0^2=0.259$ GeV$^2$, see section~\ref{Evolution}.

\subsection{$\chi$QSM}

The one-quark discrete-level wave function in the $\chi$QSM~\cite{Diakonov:2005ib} appears as the sum of a bare discrete-level wave function $F^\text{lev}_{\lambda\sigma}(\tilde k)$ and a relativistic contribution $F^\text{sea}_{\lambda\sigma}(\tilde k)$ due to the distortion of the Dirac sea
\begin{equation}
F_{\lambda\sigma}(\tilde k)=F^\text{lev}_{\lambda\sigma}(\tilde k)+F^\text{sea}_{\lambda\sigma}(\tilde k).
\end{equation}
The index $\lambda$ refers to the light-cone helicity of the bound quark and $\sigma$ to its canonical spin\footnote{In Ref.~\cite{Diakonov:2005ib} 
one writes the quark wave function as $\tilde F^{j\sigma}$, where $j$ is the quark isospin and $\sigma$ the light-cone helicity (merely referred to as ``spin''). In the model, canonical spin $s$ and isospin $j$ are coupled into grand-spin equal to zero. This means that any rotation in spin space can be compensated by a rotation in isospin space. One can then interchange canonical spin and isospin indices provided a multiplication by $\epsilon_{js}$. Our quark wave function is then related to the quark wave function $\tilde F^{j\lambda}$ of Ref.~\cite{Diakonov:2005ib} as $F_{\lambda\sigma}\equiv \sqrt{\tfrac{\uM_N}{2\pi}}\,\epsilon_{j\sigma}\tilde F^{j\lambda}$.}.

\begin{figure}[ht]
\begin{center}
\epsfig{file=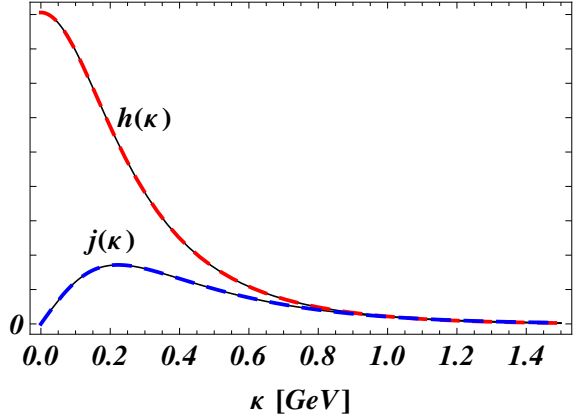,  width=.5\columnwidth}
\end{center}
\caption{\footnotesize{Comparison between the exact numerical functions $h(\kappa)$ and $j(\kappa)$ (thin solid) and their approximate form given by eq.~\eqref{approx} (thick dashed). The normalization factor is left free.}}
\label{fig13}
\end{figure}

The bare discrete-level wave function is given by
\begin{equation}\label{Flev}
F^\text{lev}_{\lambda\sigma}(\tilde k)=\left[h(\kappa)+\left(\kappa_z+i\vec\kappa_\perp\times\vec\sigma_\perp\right)\frac{j(\kappa)}{\kappa}\right]_{\lambda\sigma},
\end{equation}
where $\kappa\equiv|\vec\kappa|$ and the $z$ component of the three-vector $\vec\kappa$ is given by $\kappa_z=y\mathcal M_N-E_\text{lev}$ with the classical soliton mass $\mathcal M_N=1.207$ GeV and the energy of the discrete level $E_\text{lev}=0.2$ GeV. The functions $h(\kappa)$ and $j(\kappa)$ are the upper and lower components, in momentum space, of the Dirac spinor for a bound quark. Up to a global normalization factor, we found that they are very well approximated by the following forms
\begin{equation}\label{approx}
h(\kappa)=\frac{A_h}{1+B_h\,\kappa^2}\,e^{-C_h\,\kappa^2}\qquad\text{and}\qquad j(\kappa)=\frac{1+A_j\,\kappa+B_j\,\kappa^2}{1+C_j\,\kappa+D_j\,\kappa^2}\,\kappa\,e^{-E_j\,\kappa^{F_j}},
\end{equation}
with the parameters
\begin{align}
A_h&=0.64,&A_j&=0.72,&D_j&=12.40,\nonumber\\
B_h&=12.77,&B_j&=4.20,&E_j&=3.48,\\
C_h&=1.11,&C_j&=-1.06,&F_j&=1.17,&\nonumber
\end{align}
all in appropriate GeV units (see figure~\ref{fig13} for a comparison with the exact numerical solutions). 

\begin{figure}[t]
\begin{center}
\epsfig{file=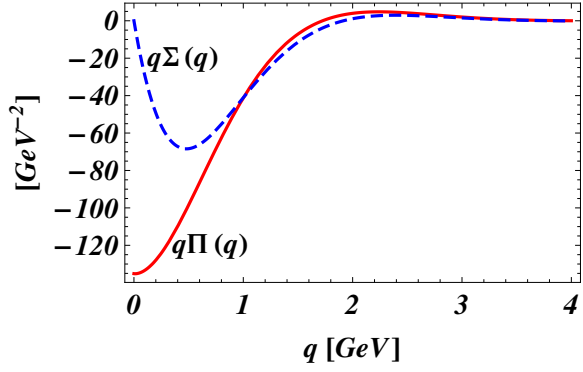,  width=.5\columnwidth}
\end{center}
\caption{\footnotesize{Scalar (dashed blue) and pseudoscalar (solid red) parts of the relativistic mean field in the baryon.}}
\label{fig14}
\end{figure}

The relativistic contribution to the discrete-level wave function due to the distortion of the Dirac sea is given by
\begin{equation}\label{Fsea}
F^\text{sea}_{\lambda\sigma}(\tilde k)=-\int\ud^3\tilde k'\,W_{\lambda\sigma,\lambda'\sigma'}(\tilde k,\tilde k')\left[\sigma_3\,h(\kappa')-\vec\kappa'\cdot\vec\sigma\,\frac{j(\kappa')}{\kappa'}\right]^{\lambda'\sigma'},
\end{equation}
where $\ud^3\tilde k'\equiv\ud y'\,\ud^2\kappa'_\perp/(2\pi)^2$. It depends linearly on the quark-antiquark pair wave function $W_{\lambda\sigma,\lambda'\sigma'}(\tilde k,\tilde k')$ whose approximate expression is
\begin{equation}
\begin{split}
W_{\lambda\sigma,\lambda'\sigma'}(\tilde k,\tilde k')&=\frac{M_Q\mathcal M_N}{2\pi Z}\left[\Sigma_{\lambda\sigma,\lambda'\sigma'}(\tilde k,\tilde k')-\Pi_{\lambda\sigma,\lambda'\sigma'}(\tilde k,\tilde k')\right],\\
\Sigma_{\lambda\sigma,\lambda'\sigma'}(\tilde k,\tilde k')&=\Sigma(q)\,\delta_{\sigma'\sigma}\left[M_Q(y'-y)\sigma_3+\vec Q_\perp\cdot\vec\sigma_\perp\right]_{\lambda\lambda'},\\
\Pi_{\lambda\sigma,\lambda'\sigma'}(\tilde k,\tilde k')&=\frac{\Pi(q)}{q}\left(\vec q\cdot\vec\sigma\right)_{\sigma'\sigma}\left[-M_Q(y+y')+i\vec Q_\perp\times\vec\sigma_\perp\right]_{\lambda\lambda'},
\end{split}
\end{equation}
where $M_Q=0.345$ GeV is the constituent quark mass and $\vec q=(\vec\kappa_\perp+\vec\kappa'_\perp,(y+y')\mathcal M_N)$ is the total momentum of the quark-antiquark pair. In order to simplify the notation we used
\begin{equation}
 \vec Q_\perp=y\vec\kappa'_\perp-y'\vec\kappa_\perp\qquad\text{and}\qquad Z=yy'\mathcal M_N\left[q_z+\bar\omega+\bar\omega'\right]
\end{equation} 
with $\bar\omega=(M^2_Q+\kappa^2_\perp)/y\mathcal M_N$ and $\bar\omega'=(M^2_Q+\kappa'^2_\perp)/y'\mathcal M_N$. The functions $\Sigma(q)$ and $\Pi(q)$, shown in figure~\ref{fig14}, correspond respectively to the scalar and pseudoscalar parts of the relativistic mean field in the baryon. 
Note that, in principle, the constituent quark mass $M_Q(p)$ in the model depends on the quark momentum $p$. This can be seen as a form factor that cuts off momenta at some characteristic scale which corresponds in the instanton picture to the inverse average size of instantons $1/\bar\rho\approx 0.6$ GeV. One then usually considers the scale of the $\chi$QSM to be about $Q_0^2=0.36$ GeV$^2$. However, since in this study we restrict ourselves to the 3Q component only, we may consider that the effective scale of our calculations is actually about $Q_0^2=0.259$ GeV$^2$ like in the LCCQM. In actual calculations, the constituent quark mass is replaced by a constant $M_Q=M_Q(0)$ and the decrease of the function $M_Q(p)$ is mimicked by the UV Pauli-Villars cutoff at $M_\text{PV}=0.557$ GeV~\cite{Diakonov:1996sr,Diakonov:1997vc}. This value has been chosen from the requirement that the pion decay constant $F_\pi=93$ MeV is reproduced from $M_Q=0.345$ GeV. The integrals in eq.~\eqref{Fsea} are convergent and do not require regularization. Nevertheless, it is not clear to us whether one should still apply the Pauli-Villars prescription or not. For this reason, when studying the effects of the relativistic contribution $F^\text{sea}$, we performed calculations both without and with Pauli-Villars prescription.

\begin{figure}[h!]
\begin{center}
\epsfig{file=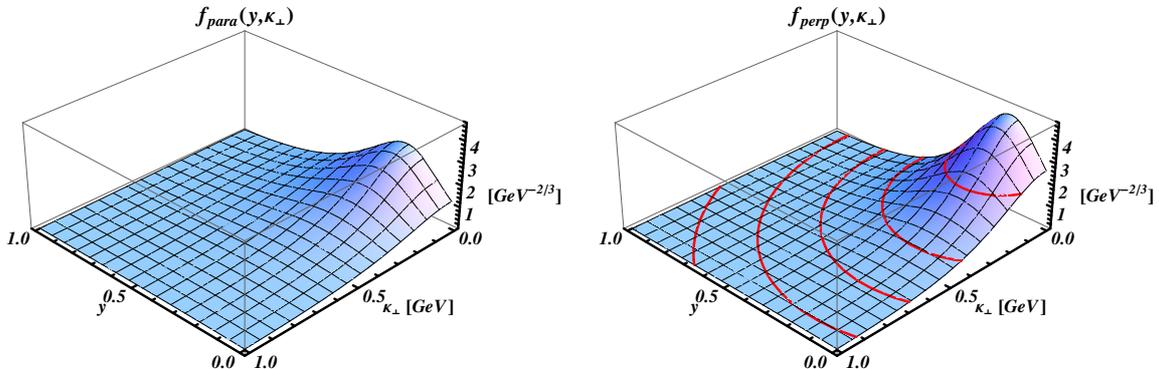,  width=1.\columnwidth}
\end{center}
\caption{\footnotesize{The two independent components $f_{/\!\!/}(y,\kappa_\perp)$ and $f_\perp(y,\kappa_\perp)$ of the bare discrete-level wave function $F^\text{lev}(\tilde k)$. They are normalized so as to have two up and one down quarks in the proton. As long as we neglect the relativistic contribution $F^\text{sea}(\tilde k)$, $f_\perp(y,\kappa_\perp)$ is actually a function of $\kappa=\sqrt{(y\mathcal M_N-E_\text{lev})^2+\kappa_\perp^2}$} only, as one can see from Eq.~\eqref{maincomp}. The thick elliptic curves represent constant values of $\kappa$.}
\label{fig15}
\end{figure}

\begin{figure}[h!]
\begin{center}
\epsfig{file=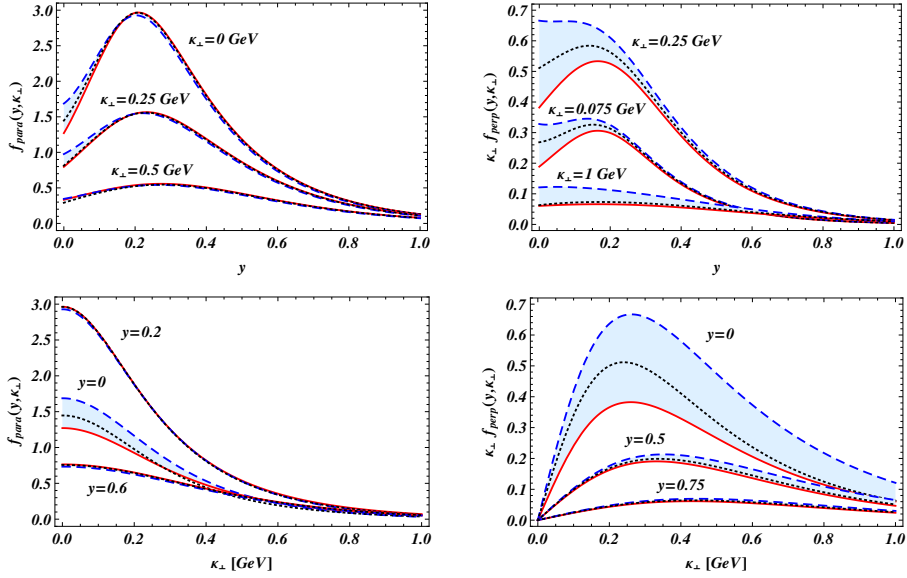,  width=0.8\columnwidth}
\end{center}
\caption{\footnotesize{Comparison between different versions of the discrete-level wave function, see text. Solid red curves: $F^\text{lev}$ only ($\chi$QSM I). Dashed blue curves: $F^\text{lev}+F^\text{sea}$ ($\chi$QSM II). Dotted black curves: $F^\text{lev}+F^\text{sea}$ with Pauli-Villars regularization ($\chi$QSM III). The light bands give the range of possible values for the discrete level, depending on the treatment of the sea contribution.
 Clearly, only the region around $y=0$ is affected in an appreciable manner by the relativistic contribution $F^\text{sea}$.}}
\label{fig16}
\end{figure}

The discrete-level wave function $F_{\lambda\sigma}(\tilde k)$ seems to be a quite complicated function of $\tilde k$, $\lambda$ and $\sigma$. Fortunately, its structure is a bit simpler than at first sight
\begin{equation}\label{Fstruct}
F_{\lambda\sigma}(\tilde k)=\begin{pmatrix}f_{/\!\!/}(y,\kappa_\perp)&\kappa_L\,f_\perp(y,\kappa_\perp)\\-\kappa_R\,f_\perp(y,\kappa_\perp)&f_{/\!\!/}(y,\kappa_\perp)\end{pmatrix},
\end{equation}
where $\kappa_{R,L}=\kappa_x\pm i\kappa_y$. The discrete-level wave function has just two independent components $f_{/\!\!/}(y,\kappa_\perp)$ and $f_\perp(y,\kappa_\perp)$ which are functions of $y$ and $\kappa_\perp$ only. If we neglect the relativistic contribution $F^\text{sea}_{\lambda\sigma}(\tilde k)$, these functions can be written as
\begin{equation}\label{maincomp}
f_{/\!\!/}(y,\kappa_\perp)=h(\kappa)+\kappa_z\,j(\kappa)/\kappa\qquad\text{and}\qquad f_\perp(y,\kappa_\perp)=j(\kappa)/\kappa.
\end{equation}
Three-dimensional plots of these functions are shown in figure~\ref{fig15}. Unless mentioned explicitly, all the results presented in this paper for the $\chi$QSM have been obtained with this wave function.
One expects the relativistic contribution $F^\text{sea}_{\lambda\sigma}(\tilde k)$ to change Eq.~\eqref{maincomp} in an appreciable manner only around $y=0$. This is confirmed by an explicit calculation of the complete discrete-level wave function $F_{\lambda\sigma}(\tilde k)$, both without and with Pauli-Villars prescription. In figure~\ref{fig16} the discrete-level wave function in three versions of $\chi$QSM are compared. 

\subsection{Generalized Melosh rotation}

There are striking similitudes at the 3Q level between the $\chi$QSM and the LCCQM. For example, both models exhibit the $SU(6)$ spin-flavor symmetry and are based on a completely symmetric momentum wave function. In the LCCQM the quarks are free while they are bound by a relativistic mean field in the $\chi$QSM. This means that in both cases the quarks are considered without mutual interactions. Usually, it is an incredibly difficult task to relate the (instant-form) canonical spin to the light-cone helicity. But since in the models we considered here there are no mutual interactions, the relation turns out to be rather simple. In the LCCQM, the quarks being free, it is well known that their canonical spin and light-cone helicity are just related by a Melosh rotation~\cite{Melosh:1974cu}, which is a special case of Wigner rotation. The amplitude and the axis of rotation depend on the quark momentum. Is it the same in the $\chi$QSM? Since here the quarks are not free, it is clear that one cannot use the Melosh rotation. In this model, it is the discrete-level wave function $F_{\lambda\sigma}(\tilde k)$ that tells us how to relate the canonical spin $\sigma$ to the light-cone helicity $\lambda$, and that this relation depends on the quark momentum $\tilde k$. Dividing $F_{\lambda\sigma}(\tilde k)$ by $f^2_{/\!\!/}+\kappa_\perp^2f^2_\perp$ in Eq.~\eqref{Fstruct}, we obtain a unitary $2\times 2$ matrix, \emph{i.e.} here also light-cone helicity and canonical spin are just related by a rotation. This explains the introduction of the generalized Melosh rotation given by Eq.~\eqref{generalizedMelosh}.

If we consider the limit of vanishing mean field in the $\chi$QSM, the quarks become free and we expect the generalized Melosh rotation to reduce to the ordinary Melosh rotation. Let us check this explicitly. The relativistic contribution $F^\text{sea}$ is directly proportional to the mean field represented by $\Sigma(q)$ and $\Pi(q)$. This means that in the limit of vanishing mean field $F^\text{sea}\to 0$. Let us then focus on $F^\text{lev}$. Since quarks becomes free, we have to replace in Eq.~\eqref{Flev} the discrete-level energy $E_\text{lev}$, the soliton mass $\mathcal M_N$ and the constituent quark mass $M_Q$ by the free quark energy $\omega$, the free invariant mass $\mathcal M_0$ and the free constituent quark mass $m$, respectively. Moreover one obtains that the upper and lower components are simply proprotional $j(\kappa)\to\kappa\,h(\kappa)/(\omega+m)$. Collecting all the parts, we find
\begin{equation}
F_{\lambda\sigma}(\tilde k)\to h(\kappa)\left[m+y\mathcal M_0+i\vec\sigma\cdot\left(\hat e_z\times\vec\kappa_\perp\right)\right]_{\lambda\sigma}=f(\tilde k)\,D^{1/2*}_{\sigma\lambda}(R_{cf}(\tilde k)),
\end{equation}
where $f(\tilde k)$ is some momentum wave function and $D^{1/2*}_{\sigma\lambda}(R_{cf}(\tilde k))$ is the matrix element of the Melosh rotation, see Eq.~\eqref{MatMelosh}. As announced, we naturally recover the Melosh rotation in the limit of vanishing mean field.

\section{LCWF overlap}\label{spinhelrot}

We derive in this appendix the master formula written in section \ref{masterformula}. Using the expression for the LCWF given by Eq.~\eqref{LCWF} for the overlap representation of the correlator tensor $W^{\mu\nu}$, we obtain Eq.~\eqref{master1} with the tensor $\mathcal A^{\mu\nu}$ given by
\begin{equation}
\mathcal A^{\mu\nu}(r',r)=\frac{1}{2}\sum_{\Lambda',\Lambda,\sigma'_i,\sigma_i}(\bar\sigma^\mu)^{\Lambda\Lambda'}\,\Phi_{\Lambda'}^{\sigma'_1\sigma'_2\sigma'_3}\,\Phi_\Lambda^{\sigma_1\sigma_2\sigma_3}\,N^\nu_{\sigma'_1\sigma_1}(\tilde k'_1,\tilde k_1)\,N^0_{\sigma'_2\sigma_2}(\tilde k'_2,\tilde k_2)\,N^0_{\sigma'_3\sigma_3}(\tilde k'_3,\tilde k_3),
\end{equation}
where $N^\nu(\tilde k',\tilde k)\equiv D^\dag(\tilde k')\,\bar\sigma^\nu\,D(\tilde k)$ are $2\times 2$ matrices. We decompose $D(\tilde k)$ and $N^\nu(\tilde k',\tilde k)$ on the basis $\sigma_\mu=\bar\sigma^\mu$ and introduce the corresponding components $d^\mu$ and $O^{\mu\nu}$
\begin{eqnarray}
D(\tilde k)&=&\sigma_\mu\,d^\mu,\qquad d^\mu=\left(\hat e_z\cdot\hat K,i\,\hat e_z\times\hat K\right),\\
N^\nu(\tilde k',\tilde k)&=&\sigma_\mu\,O^{\mu\nu},
\end{eqnarray}
where we used the notation $\hat a=\vec a/|\vec a|$. Since $\frac{1}{2}\uTr\left[\bar\sigma^\mu\sigma^\nu\right]=g^{\mu\nu}$, the components of the matrices $N^\nu(\tilde k',\tilde k)$ in the basis $\sigma_\mu$ are given by
\begin{equation}
\begin{split}
O^{\mu\nu}&=\frac{1}{2}\uTr\left[\bar\sigma^\mu D^\dag(\tilde k')\bar\sigma^\nu D(\tilde k)\right]\\
&=\frac{1}{2}\uTr\left[\bar\sigma^\mu\sigma^\alpha\bar\sigma^\nu\sigma^\beta\right]d'^*_\alpha d_\beta.
\end{split}
\end{equation} 
Now using the identity ($\epsilon_{0123}=+1$)
\begin{equation}\label{identity}
\frac{1}{2}\uTr\left[\bar\sigma^\mu\sigma^\nu\bar\sigma^\alpha\sigma^\beta\right]=g^{\mu\nu}g^{\alpha\beta}+g^{\mu\beta}g^{\alpha\nu}-g^{\mu\alpha}g^{\nu\beta}-i\,\epsilon^{\mu\nu\alpha\beta},
\end{equation}
we obtain
\begin{equation}
O^{\mu\nu}=d'^{*\mu}d^\nu+d'^{*\nu}d^\mu-\left(d'^*\cdot d\right)g^{\mu\nu}+i\,\epsilon^{\mu\nu\alpha\beta}d'^*_\alpha d_\beta,
\end{equation}
or more explicitly\footnote{Note that the elements of this matrix have already been obtained some years ago using the Melosh rotation in the LCCQM. With the notations of refs.~\cite{Boffi:2002yy,Boffi:2003yj,Pasquini:2005dk}, the matrix $O^{\mu\nu}$ reads
\begin{equation*}
O^{\mu\nu}=\frac{1}{|\vec K'||\vec K|}\begin{pmatrix}
A&-iB_{Tx}&-i\tilde A_T&-i\tilde B_z\\
iB_x&A_T&\tilde B_{Tx}&\tilde B_y\\
iB_y&B_{Tz}&\tilde B_{Ty}&\tilde B_x\\
iB_z&B_{Ty}&\tilde B_{Tz}&\tilde A
\end{pmatrix}.
\end{equation*}}
\begin{equation}
O^{\mu\nu}=\frac{1}{|\vec K'||\vec K|}\begin{pmatrix}
\vec K'\cdot\vec K&i\left(\vec K'\times\vec K\right)_x&i\left(\vec K'\times\vec K\right)_y&-i\left(\vec K'\times\vec K\right)_z\\
i\left(\vec K'\times\vec K\right)_x&\vec K'\cdot\vec K-2K'_xK_x&-K'_xK_y-K'_yK_x&K'_xK_z+K'_zK_x\\
i\left(\vec K'\times\vec K\right)_y&-K'_yK_x-K'_xK_y&\vec K'\cdot\vec K-2K'_yK_y&K'_yK_z+K'_zK_y\\
i\left(\vec K'\times\vec K\right)_z&-K'_zK_x-K'_xK_z&-K'_zK_y-K'_yK_z&-\vec K'\cdot\vec K+2K'_zK_z
\end{pmatrix}.
\end{equation}

Let us come back to $\mathcal A^{\mu\nu}$. We can now write
\begin{equation}\label{Amunu}
\mathcal A^{\mu\nu}(r',r)=\frac{1}{2}\sum_{\Lambda',\Lambda,\sigma'_i,\sigma_i}(\bar\sigma^\mu)^{\Lambda\Lambda'}\,\Phi_{\Lambda'}^{\sigma'_1\sigma'_2\sigma'_3}\,\Phi_\Lambda^{\sigma_1\sigma_2\sigma_3}\,(\sigma_\alpha)_{\sigma'_1\sigma_1}\,(\sigma_\beta)_{\sigma'_2\sigma_2}\,(\sigma_\gamma)_{\sigma'_3\sigma_3}\,O_1^{\alpha\nu}\,l_2^\beta\,l_3^\gamma,
\end{equation}
where we defined the four-component vectors $l_i^\mu=O_i^{\mu 0}$. We have to sum over all the polarizations $\Lambda',\Lambda,\sigma'_i,\sigma_i$ which seems \emph{a priori} quite involved. We can however already guess the actual form of the result by standard tensorial analysis. Indeed we must end up with a tensor with indices $\mu$ and $\nu$, and this tensor has to be constructed out of $O_1^{\alpha\nu}$, $l_2^\beta$ and $l_3^\gamma$. There are only three possible terms and $\mathcal A^{\mu\nu}$ then takes the form
\begin{equation}
\mathcal A^{\mu\nu}=A\,O_1^{\mu\nu}\left(l_2\cdot l_3\right)+B\,l_2^\mu\left(l_3\cdot O_1\right)^\nu+C\,l_3^\mu\left(l_2\cdot O_1\right)^\nu.
\end{equation}
Since spectator quarks are equivalent, we should have $B=C$. If one is not interested in the quark flavor, then all the three quarks are equivalent and we have $A=B=C$ which can be absorbed in the normalization of the LCWF. 

For the purpose of the present paper, let us determine the flavor coefficients $A$ and $B$ in the case of a proton target. Since in the process considered each individual quark flavor is conserved, it is convenient to divide the $SU(6)$ spin-flavor wave function $\Phi_\Lambda^{\sigma_1\sigma_2\sigma_3}$ into three terms
\begin{equation}
\Phi_\Lambda^{\sigma_1\sigma_2\sigma_3}=\Phi_{\Lambda,uud}^{\sigma_1\sigma_2\sigma_3}+\Phi_{\Lambda,udu}^{\sigma_1\sigma_2\sigma_3}+\Phi_{\Lambda,duu}^{\sigma_1\sigma_2\sigma_3}.
\end{equation}
Up to an overall normalization factor, $\Phi_{\Lambda,uud}^{\sigma_1\sigma_2\sigma_3}$ can be written in terms of Kronecker and Levi-Civita symbols (see \emph{e.g.} \cite{Diakonov:2005ib,Lorce:2007as})
\begin{equation}
\Phi_{\Lambda,uud}^{\sigma_1\sigma_2\sigma_3}=\delta_\Lambda^{\sigma_1}\,\epsilon^{\sigma_2\sigma_3}+\delta_\Lambda^{\sigma_2}\,\epsilon^{\sigma_1\sigma_3}.
\end{equation}
The $udu$ and $duu$ terms are obtained from the $uud$ term by permuting the labels $(2,3)$ and $(1,3)$, respectively. Using the relation $(\sigma^\mu)_{\sigma'\sigma}\,\epsilon^{\sigma'\lambda'}\,\epsilon^{\sigma\lambda}=(\bar\sigma^\mu)^{\lambda\lambda'}$
and then the identity~\eqref{identity}, we get the coefficients
\begin{equation}
\begin{tabular}{c|ccc}
&$A^p$&$B^p$&$C^p$\\\hline
$uud$&$2$&$2$&$-1$\\
$udu$&$2$&$-1$&$2$\\
$duu$&$-1$&$2$&$2$
\end{tabular}
\label{coeff}
\end{equation}
Since the first quark was chosen to be the active one, the sum of the two first lines corresponds to the contribution of the up flavor, while the last line corresponds to the contribution of the down flavor.


\begin{thebibliography}{10}

\bibitem{Meissner:2008ay}
  S.~Meissner, A.~Metz, M.~Schlegel and K.~Goeke,
  {\it Generalized parton correlation functions for a spin-0 hadron},
  {\em JHEP} {\bf 0808} (2008) 038
  [\href{http://arxiv.org/abs/0805.3165}{{\tt arXiv:0805.3165}}].

\bibitem{Meissner:2009ww}
  S.~Meissner, A.~Metz and M.~Schlegel,
  {\it Generalized parton correlation functions for a spin-1/2 hadron},
  {\em JHEP} {\bf 0908} (2009) 056
  [\href{http://arxiv.org/abs/0906.5323}{{\tt arXiv:0906.5323}}].

\bibitem{Ji:2003ak}
  X.~d.~Ji,
  {\it Viewing the proton through ``color''-filters},
  {\em Phys.\ Rev.\ Lett.}  {\bf 91} (2003) 062001
  [\href{http://arxiv.org/abs/hep-ph/0304037}{{\tt hep-ph/0304037}}].

\bibitem{Belitsky:2003nz}
  A.~V.~Belitsky, X.~d.~Ji and F.~Yuan,
  {\it Quark imaging in the proton via quantum phase-space distributions},
  {\em Phys.\ Rev.} {\bf D69} (2004) 074014
  [\href{http://arxiv.org/abs/hep-ph/0307383}{{\tt hep-ph/0307383}}].

\bibitem{Belitsky:2005qn}
  A.~V.~Belitsky and A.~V.~Radyushkin,
  {\it Unraveling hadron structure with generalized parton distributions},
  {\em Phys.\ Rept.}  {\bf 418} (2005) 1
  [\href{http://arxiv.org/abs/hep-ph/0504030}{{\tt hep-ph/0504030}}].

\bibitem{Soper:1976jc}
  D.~E.~Soper,
  {\it The parton model and the Bethe-Salpeter wave function},
  {\em Phys.\ Rev.}  {\bf D15} (1977) 1141.
  
\bibitem{Burkardt:2000za}
  M.~Burkardt,
  {\it Impact parameter dependent parton distributions and off-forward parton
  distributions for zeta $\to$ 0},
  {\em Phys.\ Rev.} {\bf D62} (2000) 071503
  [Erratum-ibid. {\bf D66} (2002) 119903]
  [\href{http://arxiv.org/abs/hep-ph/0005108}{{\tt hep-ph/0005108}}].

\bibitem{Burkardt:2002hr}
  M.~Burkardt,
 {\it Impact parameter space interpretation for generalized parton
  distributions},
  {\em Int.\ J.\ Mod.\ Phys.}   {\bf A18} (2003) 173.
  [\href{http://arxiv.org/abs/hep-ph/0207047}{{\tt hep-ph/0207047}}].

\bibitem{Diehl:2000xz}
  M.~Diehl, T.~Feldmann, R.~Jakob and P.~Kroll,
  {\it The overlap representation of skewed quark and gluon distributions},
  {\em Nucl.\ Phys. } {\bf B596} (2001) 33
  [Erratum-ibid.\   {\bf B605} (2001) 647]
  [\href{http://arxiv.org/abs/hep-ph/0009255}{{\tt hep-ph/0009255}}].

\bibitem{Brodsky:2000xy}
  S.~J.~Brodsky, M.~Diehl and D.~S.~Hwang,
  {\it Light-cone wavefunction representation of deeply virtual Compton
  scattering},
  {\em Nucl.\ Phys.}   {\bf B596} (2001) 99
  [\href{http://arxiv.org/abs/hep-ph/0009254}{{\tt hep-ph/0009254}}].

\bibitem{Diehl:1999tr}
  M.~Diehl, T.~Feldmann, R.~Jakob and P.~Kroll,
  {\it Skewed parton distributions in real and virtual Compton scattering},
  {\em Phys.\ Lett.  } {\bf B460} (1999) 204
  [\href{http://arxiv.org/abs/hep-ph/9903268}{{\tt hep-ph/9903268}}].

\bibitem{Boffi:2007yc}
  S.~Boffi and B.~Pasquini,
  {\it Generalized parton distributions and the structure of the nucleon},
  {\em Riv.\ Nuovo Cim.}  {\bf 30} (2007) 387
  [\href{http://arxiv.org/abs/0711.2625}{{\tt arXiv:0711.2625}}].

\bibitem{Boffi:2002yy}
  S.~Boffi, B.~Pasquini and M.~Traini,
  {\it Linking generalized parton distributions to constituent quark models},
  {\em Nucl.\ Phys.}   {\bf B649} (2003) 243
  [\href{http://arxiv.org/abs/hep-ph/0207340}{{\tt hep-ph/0207340}}].

\bibitem{Boffi:2003yj}
  S.~Boffi, B.~Pasquini and M.~Traini,
  {\it Helicity-dependent generalized parton distributions in constituent  quark
  models},
  {\em Nucl.\ Phys.}   {\bf B680} (2004) 147
  [\href{http://arxiv.org/abs/hep-ph/0311016}{{\tt hep-ph/0311016}}].

\bibitem{Pasquini:2005dk}
  B.~Pasquini, M.~Pincetti and S.~Boffi,
  {\it Chiral-odd generalized parton distributions in constituent quark  models},
  {\em Phys.\ Rev.} {\bf D72} (2005) 094029
  [\href{http://arxiv.org/abs/hep-ph/0510376}{{\tt hep-ph/0510376}}].

\bibitem{Diakonov:1984vw}
  D.~Diakonov and V.~Y.~Petrov,
  {\it Chiral condensate in the instanton vacuum},
  {\em Phys.\ Lett.} {\bf B147} (1984) 351.
  
\bibitem{Diakonov:1985eg}
  D.~Diakonov and V.~Y.~Petrov,
  {\it A theory of light quarks in the instanton vacuum},
  {\em Nucl.\ Phys.} {\bf B272} (1986) 457.

\bibitem{Diakonov:1986yh}
	D.~Diakonov and V.~Y.~Petrov,
  {\it Chiral theory of nucleons},
  {\em JETP Lett.}  {\bf 43} (1986) 75
  [Pisma Zh.\ Eksp.\ Teor.\ Fiz.\  {\bf 43} (1986) 57].

\bibitem{Diakonov:1987ty}
  D.~Diakonov, V.~Y.~Petrov and P.~V.~Pobylitsa,
  {\it A chiral theory of nucleons},
  {\em Nucl.\ Phys.} {\bf B306} (1988) 809.
  
\bibitem{Petrov:2002jr}
  V.~Y.~Petrov and M.~V.~Polyakov,
  {\it Light cone nucleon wave function in the quark soliton model},
  \href{http://arxiv.org/abs/hep-ph/0307077}{{\tt hep-ph/0307077}}.

\bibitem{Diakonov:2005ib}
  D.~Diakonov and V.~Petrov,
  {\it Estimate of the $\Theta^+$ width in the relativistic mean field
  approximation},
  {\em Phys.\ Rev.}   {\bf D72} (2005) 074009
  [\href{http://arxiv.org/abs/hep-ph/0505201}{{\tt hep-ph/0505201}}].

\bibitem{Lorce:2006nq}
  C.~Lorcé,
  {\it Improvement of the $\Theta^+$ width estimation method on the light cone},
  {\em Phys.\ Rev.} {\bf D74} (2006) 054019
  [\href{http://arxiv.org/abs/hep-ph/0603231}{{\tt hep-ph/0603231}}].

\bibitem{Lorce:2007as}
  C.~Lorcé,
  {\it Baryon vector and axial content up to the 7Q component},
  {\em Phys.\ Rev.} {\bf D78} (2008) 034001
  [\href{http://arxiv.org/abs/0708.3139}{{\tt arXiv:0708.3139}}].

\bibitem{Lorce:2007fa}
  C.~Lorcé,
  {\it Tensor charges of light baryons in the Infinite Momentum Frame},
  {\em Phys.\ Rev.} {\bf D79} (2009) 074027
  [\href{http://arxiv.org/abs/0708.4168}{{\tt arXiv:0708.4168}}].

\bibitem{Melosh:1974cu}
  H.~J.~Melosh,
  {\it Quarks: Currents and constituents},
  {\em Phys.\ Rev.} {\bf D9} (1974) 1095.

\bibitem{Kogut:1969xa}
  J.~B.~Kogut and D.~E.~Soper,
  {\it Quantum Electrodynamics In The Infinite Momentum Frame},
  {\em Phys.\ Rev.} {\bf D1} (1970) 2901.

\bibitem{Pasquini:2008ax}
  B.~Pasquini, S.~Cazzaniga and S.~Boffi,
  {\it Transverse momentum dependent parton distributions in a light-cone quark
  model},
  {\em Phys.\ Rev.}   {\bf D78} (2008) 034025
  [\href{http://arxiv.org/abs/0806.2298}{{\tt arXiv:0806.2298}}].

\bibitem{Avakian:2010br}
  H.~Avakian, A.~V.~Efremov, P.~Schweitzer and F.~Yuan,
  {\it The transverse momentum dependent distribution functions in the bag
  model},
  {\em Phys.\ Rev.} {\bf D81} (2010) 074035
  [\href{http://arxiv.org/abs/1001.5467}{{\tt arXiv:1001.5467}}].
  
\bibitem{Avakian:2008dz}
  H.~Avakian, A.~V.~Efremov, P.~Schweitzer and F.~Yuan,
  {\it Transverse momentum dependent distribution function $h_{1T}^\perp$ and the
  single spin asymmetry $A_{UT}^{\sin(3\phi-\phi_S)}$},
  {\em Phys.\ Rev.} {\bf D78} (2008) 114024
  [\href{http://arxiv.org/abs/0805.3355}{{\tt arXiv:0805.3355}}].

\bibitem{Pasquini:2010pa}
  B.~Pasquini and C.~Lorcé,
  {\it Modeling the transverse momentum dependent parton distributions},
  \href{http://arxiv.org/abs/1008.0945}{{\tt arXiv:1008.0945}}.

\bibitem{Musch:2010ka}
  B.~U.~Musch, P.~H\"agler, J.~W.~Negele and A.~Sch\"afer,
  {\it Exploring quark transverse momentum distributions with lattice QCD},
  \href{http://arxiv.org/abs/1011.1213}{{\tt arXiv:1011.1213}}.
  
\bibitem{Hagler:2009mb}
  Ph.~H\"agler, B.~U.~Musch, J.~W.~Negele and A.~Sch\"afer,
  {\it Intrinsic quark transverse momentum in the nucleon from lattice QCD},
  {\em Europhys.\ Lett.}  {\bf 88} (2009) 61001
  [\href{http://arxiv.org/abs/0908.1283}{{\tt arXiv:0908.1283}}].

\bibitem{Wakamatsu:2009fn}
  M.~Wakamatsu,
  {\it Transverse momentum distributions of quarks in the nucleon from the Chiral
  Quark Soliton Model},
  {\em Phys.\ Rev.} {\bf D79} (2009) 094028
  [\href{http://arxiv.org/abs/0903.1886}{{\tt arXiv:0903.1886}}].

\bibitem{Diehl:2005jf}
  M.~Diehl and Ph.~H\"agler,
  {\it Spin densities in the transverse plane and generalized transversity
  distributions},
  {\em Eur.\ Phys.\ J.}   {\bf C44} (2005) 87
  [\href{http://arxiv.org/abs/hep-ph/0504175}{{\tt hep-ph/0504175}}].

\bibitem{Burkardt:2003uw}
  M.~Burkardt,
  {\it Chromodynamic lensing and transverse single spin asymmetries},
  {\em Nucl.\ Phys.} {\bf A735} (2004) 185
  [\href{http://arxiv.org/abs/hep-ph/0302144}{{\tt hep-ph/0302144}}].

\bibitem{Burkardt:2003je}
  M.~Burkardt and D.~S.~Hwang,
  {\it Sivers asymmetry and generalized parton distributions in impact  parameter
  space},
  {\em Phys.\ Rev.}   {\bf D69} (2004) 074032
  [\href{http://arxiv.org/abs/hep-ph/0309072}{{\tt hep-ph/0309072}}].

\bibitem{Meissner:2007rx}
  S.~Meissner, A.~Metz and K.~Goeke,
  {\it Relations between generalized and transverse momentum dependent parton
  distributions},
  {\em Phys.\ Rev.} {\bf D76} (2007) 034002
  [\href{http://arxiv.org/abs/hep-ph/0703176}{{\tt hep-ph/0703176}}].
  
\bibitem{Ossmann:2004bp}
  J.~Ossmann, M.~V.~Polyakov, P.~Schweitzer, D.~Urbano and K.~Goeke,
  {\it The generalized parton distribution function $(E^u+E^d)(x,\xi,t)$ of the
  nucleon in the chiral quark soliton model},
  {\em Phys.\ Rev.} {\bf D71} (2005) 034011
  [\href{http://arxiv.org/abs/hep-ph/0411172}{{\tt hep-ph/0411172}}].

\bibitem{Polyakov:1999gs}
  M.~V.~Polyakov and C.~Weiss,
  {\it Skewed and double distributions in pion and nucleon},
  {\em Phys.\ Rev.} {\bf D60} (1999) 114017
  [\href{http://arxiv.org/abs/hep-ph/9902451}{{\tt hep-ph/9902451}}].
  
\bibitem{Penttinen:1999th}
  M.~Penttinen, M.~V.~Polyakov and K.~Goeke,
  {\it Helicity skewed quark distributions of the nucleon and chiral symmetry},
  {\em Phys.\ Rev.} {\bf D62} (2000) 014024
  [\href{http://arxiv.org/abs/hep-ph/9909489}{{\tt hep-ph/9909489}}].
  
\bibitem{Goeke:2001tz}
  K.~Goeke, M.~V.~Polyakov and M.~Vanderhaeghen,
  {\it Hard Exclusive Reactions and the Structure of Hadrons},
  {\em Prog.\ Part.\ Nucl.\ Phys.}  {\bf 47} (2001) 401
  [\href{http://arxiv.org/abs/hep-ph/0106012}{{\tt hep-ph/0106012}}].

\bibitem{Wakamatsu:2008ki}
  M.~Wakamatsu,
  {\it Chiral-odd GPDs, transversity decomposition of angular momentum, and tensor
  charges of the nucleon},
  {\em Phys.\ Rev.} {\bf D79} (2009) 014033
  [\href{http://arxiv.org/abs/0811.4196}{{\tt arXiv:0811.4196 }}].
  
\bibitem{Wakamatsu:2006dy}
  M.~Wakamatsu and Y.~Nakakoji,
  {\it Generalized form factors, generalized parton distributions and the spin
  contents of the nucleon},
  {\em Phys.\ Rev.} {\bf D74} (2006) 054006
  [\href{http://arxiv.org/abs/hep-ph/0605279}{{\tt hep-ph/0605279 }}].
  
\bibitem{Wakamatsu:2005vk}
  M.~Wakamatsu and H.~Tsujimoto,
  {\it The generalized parton distribution functions and the nucleon spin sum
  rules in the chiral quark soliton model},
  {\em Phys.\ Rev.}   {\bf D71} (2005) 074001
  [\href{http://arxiv.org/abs/hep-ph/0502030}{{\tt hep-ph/0502030 }}].

\bibitem{Martin:2009iq}     
	A.~D.~Martin, W.~J.~Stirling, R.~S.~Thorne and G.~Watt,
	{\it Parton distributions for the LHC},
	{\em Eur.\ Phys.\ J.} {\bf C63} (2009) 189
  [\href{http://arxiv.org/abs/0901.0002}{{\tt arXiv:0901.0002 }}].

\bibitem{Leader:2006xc}
  E.~Leader, A.~V.~Sidorov and D.~B.~Stamenov,
  {\it Impact of CLAS and COMPASS data on Polarized Parton Densities and Higher
  Twist},
  {\em Phys.\ Rev.} {\bf D75} (2007) 074027
  [\href{http://arxiv.org/abs/hep-ph/0612360}{{\tt hep-ph/0612360 }}].
\bibitem{Pasquini:2011tk}
  B.~Pasquini and P.~Schweitzer,
  arXiv:1103.5977 [hep-ph].


\bibitem{Anselmino:2007fs}
  M.~Anselmino, M.~Boglione, U.~D'Alesio, A.~Kotzinian, F.~Murgia, A.~Prokudin and C.~Turk,
  {\it Transversity and Collins functions from SIDIS and $e^+ e^-$ data},
  {\em Phys.\ Rev.} {\bf D75} (2007) 054032
  [\href{http://arxiv.org/abs/hep-ph/0701006}{{\tt hep-ph/0701006 }}].

\bibitem{Anselmino:2008jk}
  M.~Anselmino, M.~Boglione, U.~D'Alesio, A.~Kotzinian, F.~Murgia, A.~Prokudin and S.~Melis,
  {\it Update on transversity and Collins functions from SIDIS and $e^+ e^-$ data},
  {\em Nucl.\ Phys.\ Proc.\ Suppl.}  {\bf 191} (2009) 98
  [\href{http://arxiv.org/abs/0812.4366}{{\tt arXiv:0812.4366 }}].

\bibitem{Soffer:1994ww}
  J.~Soffer,
  {\it Positivity constraints for spin dependent parton distributions},
  {\em Phys.\ Rev.\ Lett.}  {\bf 74} (1995) 1292
  [\href{http://arxiv.org/abs/hep-ph/9409254}{{\tt hep-ph/9409254 }}].

\bibitem{Boffi:2009sh}
  S.~Boffi, A.~V.~Efremov, B.~Pasquini and P.~Schweitzer,
  {\it Azimuthal spin asymmetries in light-cone constituent quark models},
  {\em Phys.\ Rev.} {\bf D79} (2009) 094012
  [\href{http://arxiv.org/abs/0903.1271}{{\tt arXiv:0903.1271 }}].

\bibitem{Diefenthaler:2005gx}
  M.~Diefenthaler  [HERMES Collaboration],
  {\it Transversity measurements at HERMES},
  {\em AIP Conf.\ Proc.}  {\bf 792} (2005) 933
  [\href{http://arxiv.org/abs/hep-ex/0507013}{{\tt hep-ex/0507013 }}].

\bibitem{Alekseev:2008dn}
  M.~Alekseev {\it et al.}  [COMPASS Collaboration],
  {\it Collins and Sivers Transverse Spin Asymmetries for Pions and Kaons on
  Deuterons},
  \href{http://arxiv.org/abs/0802.2160}{{\tt arXiv:0802.2160 }}.
\bibitem{Efremov:2006qm}
  A.~V.~Efremov, K.~Goeke and P.~Schweitzer,
  Phys.\ Rev.\  D {\bf 73}, 094025 (2006).
\bibitem{Vogelsang:2005cs}
  W.~Vogelsang and F.~Yuan,
  Phys.\ Rev.\  D {\bf 72}, 054028 (2005).
\bibitem{Boer:2001he}
  D.~Boer,
  Nucl.\ Phys.\  B {\bf 603}, 195 (2001).
\bibitem{Abe:2005zx}
  K.~Abe {\it et al.}  [Belle Collaboration],
  Phys.\ Rev.\ Lett.\  {\bf 96} (2006) 232002.




\bibitem{Goeke:2000wv}
  K.~Goeke, P.~V.~Pobylitsa, M.~V.~Polyakov, P.~Schweitzer and D.~Urbano,
  {\it Quark distribution functions in the chiral quark-soliton model:
  Cancellation of quantum anomalies},
  {\em Acta Phys.\ Polon. {\bf B32}, (2001) 1201}
  [\href{http://arxiv.org/abs/hep-ph/0001272}{{\tt hep-ph/0001272 }}].

\bibitem{Schweitzer:2001sr}
  P.~Schweitzer, D.~Urbano, M.~V.~Polyakov, C.~Weiss, P.~V.~Pobylitsa
  and K.~Goeke,
  {\it Transversity distributions in the nucleon in the large-N$_c$ limit},
  {\em Phys.\ Rev.} {\bf D64} (2001) 034013
  [\href{http://arxiv.org/abs/hep-ph/0101300}{{\tt hep-ph/0101300 }}].

\bibitem{Diehl:2001pm}
  M.~Diehl,
  {\it Generalized parton distributions with helicity flip},
  {\em Eur.\ Phys.\ J.} {\bf C19} (2001) 485
  [\href{http://arxiv.org/abs/hep-ph/0101335}{{\tt hep-ph/0101335 }}].

\bibitem{Perdrisat:2006hj}
  C.~F.~Perdrisat, V.~Punjabi and M.~Vanderhaeghen,
  {\it Nucleon electromagnetic form factors},
  {\em Prog.\ Part.\ Nucl.\ Phys.}  {\bf 59} (2007) 694
  [\href{http://arxiv.org/abs/hep-ph/0612014}{{\tt hep-ph/0612014 }}].

\bibitem{Pasquini:2007iz}
  B.~Pasquini and S.~Boffi,
  {\it Electroweak structure of the nucleon, meson cloud and light-cone
  wavefunctions},
  {\em Phys.\ Rev.\  } {\bf D76} (2007) 074011
  [\href{http://arxiv.org/abs/0707.2897}{{\tt arXiv:0707.2897}}].

\bibitem{Pasquini:2007xz}
  B.~Pasquini and S.~Boffi,
  {\it Nucleon spin densities in a light-front constituent quark model},
  {\em Phys.\ Lett.} {\bf B653} (2007) 23
  [\href{http://arxiv.org/abs/0705.4345}{{\tt arXiv:0705.4345}}].

\bibitem{Pasquini:2006iv}
  B.~Pasquini, M.~Pincetti and S.~Boffi,
  {\it Drell-Yan processes, transversity and light-cone wavefunctions},
  {\em Phys.\ Rev.} {\bf D76} (2007) 034020
  [\href{http://arxiv.org/abs/hep-ph/0612094}{{\tt hep-ph/0612094 }}].

\bibitem{Airapetian:2007mh}
  A.~Airapetian {\it et al.}  [HERMES Collaboration],
  {\it Precise determination of the spin structure function $g_1$ of the  proton,
  deuteron and neutron},
  {\em Phys.\ Rev.\ }  {\bf D75} (2007) 012007
  [\href{http://arxiv.org/abs/hep-ex/0609039}{{\tt hep-ex/0609039 }}].

\bibitem{Nakamura:2010zzi}
  K.~Nakamura {\it et al.}  [Particle Data Group],
  {\it Review of particle physics},
  {\em J.\ Phys.} {\bf G37} (2010) 075021.
\bibitem{deFlorian:2009vb}
  D.~de Florian, R.~Sassot, M.~Stratmann and W.~Vogelsang,
  {\it Extraction of Spin-Dependent Parton Densities and Their Uncertainties},
  {\em Phys.\ Rev.} {\bf D80} (2009) 034030
  [\href{http://arxiv.org/abs/0904.3821}{{\tt arXiv:0904.3821 }}].


\bibitem{Dziembowski:1987zp}
  Z.~Dziembowski,
  {\it Relativistic model of nucleon and pion structure: static properties
and electromagnetic soft form-factors},
  {\em Phys.\ Rev.} {\bf D37} (1988) 778.
  
\bibitem{Chung:1991st}
  P.~L.~Chung and F.~Coester,
  {\it Relativistic constituent quark model of nucleon form-factors},
  {\em Phys.\ Rev.} {\bf D44} (1991) 229.

\bibitem{Cardarelli:1995dc}
  F.~Cardarelli, E.~Pace, G.~Salme and S.~Simula,
  {\it Nucleon and pion electromagnetic form-factors in a light front constituent
  quark model},
  {\em Phys.\ Lett.} {\bf B357} (1995) 267
  [\href{http://arxiv.org/abs/nucl-th/9507037}{{\tt nucl-th/9507037 }}].

\bibitem{Ma:2002ir}
  B.~Q.~Ma, D.~Qing and I.~Schmidt,
  {\it Electromagnetic form factors of nucleons in a light-cone diquark model},
  {\em Phys.\ Rev.} {\bf C65} (2002) 035205
  [\href{http://arxiv.org/abs/hep-ph/0202015}{{\tt hep-ph/0202015}}].

\bibitem{Ledwig:2010tu}
  T.~Ledwig, A.~Silva and H.~C.~Kim,
  {\it Tensor charges and form factors of SU(3) baryons in the self-consistent
  SU(3) chiral quark-soliton model},
  {\em Phys.\ Rev.} {\bf D82} (2010) 034022
  [\href{http://arxiv.org/abs/1004.3612}{{\tt arXiv:1004.3612}}].

\bibitem{Pasquini:2010af}
  B.~Pasquini and F.~Yuan,
  {\it Sivers and Boer-Mulders functions in Light-Cone Quark Models},
  {\em Phys.\ Rev.} {\bf D81} (2010) 114013
  [\href{http://arxiv.org/abs/1001.5398}{{\tt arXiv:1001.5398}}].

\bibitem{Gockeler:2006zu}
  M.~Gockeler {\it et al.}  [QCDSF Collaboration and UKQCD Collaboration],
  {\it Transverse spin structure of the nucleon from lattice QCD simulations},
  {\em Phys.\ Rev.\ Lett.}  {\bf 98} (2007) 222001
  [\href{http://arxiv.org/abs/hep-lat/0612032}{{\tt hep-lat/0612032  }}].
  
\bibitem{Ledwig:2010zq}
  T.~Ledwig, A.~Silva and H.~C.~Kim,
  {\it Anomalous tensor magnetic moments and form factors of the proton in the
  self-consistent chiral quark-soliton model},
  {\em Phys.\ Rev.} {\bf D82} (2010) 054014
  [\href{http://arxiv.org/abs/1007.1355}{{\tt arXiv:1007.1355}}].
    
\bibitem{Schlumpf:1992ce}
  F.~Schlumpf,
  {\it Relativistic constituent quark model for baryons},
  \href{http://arxiv.org/abs/hep-ph/9211255}{{\tt hep-ph/9211255}}.

\bibitem{Diakonov:1996sr}
  D.~Diakonov, V.~Petrov, P.~Pobylitsa, M.~V.~Polyakov and C.~Weiss,
  {\it Nucleon parton distributions at low normalization point in the large  
    N$_c$
  limit},
  {\em Nucl.\ Phys.} {\bf B480} (1996) 341
  [\href{http://arxiv.org/abs/hep-ph/9606314}{{\tt hep-ph/9606314}}].

\bibitem{Diakonov:1997vc}
  D.~Diakonov, V.~Y.~Petrov, P.~V.~Pobylitsa, M.~V.~Polyakov and C.~Weiss,
  {\it Unpolarized and polarized quark distributions in the large-N$_c$ limit},
  {\em Phys.\ Rev.}  {\bf D56} (1997) 4069
  [\href{http://arxiv.org/abs/hep-ph/9703420}{{\tt hep-ph/9703420}}].


\end{thebibliography}

\end{document}